\documentstyle[12pt]{article}

\makeatletter
\def\eqnarray{\stepcounter{equation}\let\@currentlabel=\theequation
\global\@eqnswtrue
\global\@eqcnt\z@\tabskip\@centering\let\\=\@eqncr
$$\halign to \displaywidth\bgroup\@eqnsel\hskip\@centering
  $\displaystyle\tabskip\z@{##}$&\global\@eqcnt\@ne
  \hfil$\displaystyle{\hbox{}##\hbox{}}$\hfil
  &\global\@eqcnt\tw@ $\displaystyle\tabskip\z@
  {##}$\hfil\tabskip\@centering&\llap{##}\tabskip\z@\cr}
\def\@sect#1#2#3#4#5#6[#7]#8{\ifnum #2>\c@secnumdepth
    \def\@svsec{}\else
    \refstepcounter{#1}\edef\@svsec{\csname the#1\endcsname.\hskip 1em
    }\fi
    \@tempskipa #5\relax
    \ifdim \@tempskipa>\z@
    \begingroup #6\relax
    \@hangfrom{\hskip #3\relax\@svsec}{\interlinepenalty \@M #8\par}
    \endgroup
    \csname #1mark\endcsname{#7}\addcontentsline
    {toc}{#1}{\ifnum #2>\c@secnumdepth \else
     \protect\numberline{\csname the#1\endcsname}\fi
           #7}\else
    \def\@svsechd{#6\hskip #3\@svsec #8\csname #1mark\endcsname
          {#7}\addcontentsline
          {toc}{#1}{\ifnum #2>\c@secnumdepth \else
     \protect\numberline{\csname the#1\endcsname}\fi
           #7}}\fi
     \@xsect{#5}}
\def\label#1{\@bsphack\if@filesw {\let\thepage\relax
   \xdef\@gtempa{\write\@auxout{\string
   \newlabel{#1}{{\thesection.\@currentlabel}{\thepage}}}}}\@gtempa
   \if@nobreak \ifvmode\nobreak\fi\fi\fi\@esphack}
\def\@eqnnum{(\thesection.\theequation)}
\def\section{\setcounter{equation}{0} \@startsection {section}{1}{\z@}
{-3.5ex
   plus -1ex minus -.2ex}{2.3ex plus .2ex}{\Large\bf}}
\newcount\@minsofar
\newcount\@min
\newcount\@cite@temp
\def\@citex[#1]#2{%
\if@filesw \immediate \write \@auxout {\string \citation {#2}}\fi
\@tempcntb\m@ne \let\@h@ld\relax \def\@citea{}%
\@min\m@ne%
\@cite{%
  \@for \@citeb:=#2\do {\@ifundefined {b@\@citeb}%
    {\@h@ld\@citea\@tempcntb\m@ne{\bf ?}%
    \@warning {Citation `\@citeb ' on page \thepage \space
    undefined}}%
{\@minsofar\z@ \@for \@scan@cites:=#2\do {%
  \@ifundefined{b@\@scan@cites}%
    {\@cite@temp\m@ne}
    {\@cite@temp\number\csname b@\@scan@cites \endcsname \relax}%
\ifnum\@cite@temp > \@min
    \ifnum\@minsofar = \z@
      \@minsofar\number\@cite@temp
      \edef\@scan@copy{\@scan@cites}\else
    \ifnum\@cite@temp < \@minsofar
      \@minsofar\number\@cite@temp
      \edef\@scan@copy{\@scan@cites}\fi\fi\fi}\@tempcnta\@min
  \ifnum\@minsofar > \z@ 
    \advance\@tempcnta\@ne
    \@min\@minsofar
    \ifnum\@tempcnta=\@minsofar 
    to it
      \ifx\@h@ld\relax
        \edef \@h@ld{\@citea\csname b@\@scan@copy\endcsname}%
    \else \edef\@h@ld{\ifmmode{-}\else--\fi\csname
    b@\@scan@copy\endcsname}%
      \fi
    \else \@h@ld\@citea\csname b@\@scan@copy\endcsname
          \let\@h@ld\relax
  \fi 
\fi}%
\def\@citea{,\penalty\@highpenalty\,}}\@h@ld}{#1}}
\def\appendixname{Appendix}
\def\appendix{\par
  \def\pre@section{\appendixname{}}
  \setcounter{section}{1}
  \@addtoreset{equation}{section}
  \def\thesection{\Alph{section}}
  \def\theequation{\arabic{equation}}}

\def\appendix{\par
  \def\pre@section{\appendixname{}}
  \setcounter{section}{1}
  \@addtoreset{equation}{section}
  \def\thesection{\Alph{section}}
  \def\theequation{\arabic{equation}}}

\makeatother

\def\b{\beta}
\def\d{\delta}

\def\a{\alpha}
\def\s{\sigma}

\def\l{\lambda}
\def\e{\epsilon}
\def\r{\rho}

\def\ds{\displaystyle}
\def\be{\begin{equation}}
\def\ee{\end{equation}}
\def\beq{\begin{eqnarray}}
\def\eeq{\end{eqnarray}}

\begin{document}
\begin{center}
\bf {Evaluation of Integrals  Representing  Correlations in XXX Heisenberg Spin Chain}
\end{center}
\phantom{a}

\vspace{1.5cm}

\centerline{H.E. Boos\footnote{
E-mail: boos@mx.ihep.su}}
\centerline{
Institute for High Energy Physics}
\centerline{ Protvino, 142284, Russia}

\phantom{a}

\vspace{0.5cm}

\phantom{a}

\centerline{V.E. Korepin \footnote{E-mail:
korepin@insti.physics.sunysb.edu}}

\centerline{C.N.~Yang Institute for Theoretical Physics}
\centerline{State University of New York at Stony Brook}
\centerline{Stony Brook, NY 11794--3840, USA}

\vspace{1.5cm}

\vskip2em
\begin{abstract}
\noindent
 We study XXX  Heisenberg spin 1/2 anti-ferromagnet.
We evaluate a  probability of formation of a
ferromagnetic string in the anti-ferromagnetic ground state in
thermodynamics limit.
We prove that  for short strings the probability can be expressed
in terms of Riemann zeta function with odd arguments.

\end{abstract}

\newpage

\section{Introduction}

 Riemann zeta function for $Re (s) > 1 $ can be defined as follows:

\be
\zeta (s) = \sum_{n=1}^{\infty}\frac{1}{n^s}
\label{zeta}
\ee

It also can be represented as a product with respect to all prime
numbers $p$
\be
\zeta (s) = \prod_{p}  ( 1-p^{-s}  ) ^{-1}
\ee

It can be analytically continued in the whole complex plane of $s$.
It has only one pole, at $s=1$. Riemann zeta function satisfies a 
functional equation
\be
\zeta (s) = 2^s \pi^{s-1} \sin ( {s\pi /2}) \Gamma (1-s) \zeta (1-s)
\ee

It has 'trivial' zeros at $s=-2n$ (
$n>1$ is an integer).
The famous Riemann hypothesis  \cite{R}  states that nontrivial zeros
belong to the straight line
$Re (s) = 1/2$. Recently Montgomery and Odlyzko   conjectured that for large values of 
imaginary part of $s$ the distribution of zeros can be described by GUE of random matrices, see 
 \cite{mon} and \cite{odl}.
Forrester and Odlyzko  related the   problem of distribution of zeros  to Painleve
 differential  equation and integrable integral operators \cite{fo}.
Riemann zeta function is useful for study of distribution of prime
numbers on the
real axis  \cite{TIT}. The values of Riemann zeta function at special
points were studied in
\cite{za} , \cite{bbbl} .
At even values of its argument  zeta function can be expressed in
 terms of powers of $\pi$ and Bernoulli's numbers
\be
\zeta (2n) = (-1)^{n+1} 2^{2n-1} \pi ^{2n} B_{2n} /(2n)! 
\label{zeta2n}
\ee
The values of Riemann zeta function at odd arguments
provide infinitely many different irrational numbers \cite{tr} .
Riemann zeta function  plays an important role, not only in pure
mathematics
but also  theoretical physics. Some Feynman diagrams in
quantum field theory can be expressed in terms of $\zeta (n)$,
see, for example, \cite{KR} .
In statistical mechanics Riemann zeta function  was used for the
description of chaotic systems \cite{kna}.
One can find more information and citation  on  the following web-cite
{http://www.maths.ex.ac.uk/~mwatkins/} .

We argue that  $\zeta (n)$ is also important for exactly solvable
models.
The most famous integrable models is the Heisenberg  XXX spin chain.
This model was first suggested by Heisenberg  \cite{Heis} in 1928
and
solved by Bethe  \cite{B} in 1931 .
Since that time it found multiple applications in solid state physics
and statistical mechanics.

The Hamiltonian of the XXX spin chain can be written like this

\be
H = \sum_{i=1}^N \>
(\s^x_i\s^x_{i+1}\; + \;\s^y_i\s^y_{i+1}\; +\; \s^z_i\s^z_{i+1}\;-1\;)
\label{H}
\ee
Here $N$ is the length of the lattice and  $\s^x_i,\s^y_i,\s^z_i$ are
Pauli  matrices. We consider thermodynamics limit , when $N$ goes to infinity.
 The sign in front of the Hamiltonian indicates that we
are considering the
anti-ferromagnetic case. We consider periodic boundary conditions.
Notice that this Hamiltonian  annihilates the ferromagnetic state [
all spins up].

The construction of the anti-ferromagnetic ground state wave
function $|AFM>$  can be credited to
Hulth\'{e}n  \cite{H}. An important
correlation function was defined in \cite{KIEU}.
It was called the emptiness formation probability

$$
P(n) = <AFM|\prod_{j=1}^n P_j|AFM>
$$
where $P_j= (1+\s^z_j)/2$ is a projector on the state with spin up in
$j$th lattice site.
Averaging is over the
anti-ferromagnetic ground state. It describes the
probability of formation of a
ferromagnetic string of the length $n$ in the
anti-ferromagnetic background  $|AFM>$ .
In this paper we shall first study short strings ( $n$ is small),
in the end we shall discuss
long distance asymptotic ( at finite temperature).
The four first values of the emptiness-formation probability look as
follows:
\beq
&{\ds P(1)\;=\;{1\over 2}}\;=\;0.5,&\label{P1}\\
&{\ds P(2)\;=\;{1\over 3} (1\; -\; \ln{2})}\;=\;0.102284273,&
\label{P2}\\
&{\ds P(3)\;=\;{1\over 4}\; - \;\ln{2}\;+\;{3\over 8}\;\zeta(3)}\;=
\;0.007624158,&
\label{P3}\\
&{\ds P(4)\;=\;{1\over 5}\; -\; 2\ln{2} \; + \; {173\over 60}
\,\zeta(3)
\; -\; {11\over 6} \,\zeta(3)\, \ln{2}\;-\; {51\over 80} \, \zeta^2(3)
\;
}&
\nonumber\\
&{\ds
 -\; {55\over 24}\,\zeta(5)
\; + \; {85\over 24}\,\zeta(5)\,  \ln{2}\;=\;0.000206270}&
\label{P4}
\eeq
 Let us comment.
The value of  $P(1)$ is evident from the symmetry, $P(2)$
can be extracted from the explicit expression of the ground
state energy \cite{H}.
$P(3)$  can be extracted from the results of M.Takahashi \cite{T1}
on the calculation of the nearest neighbor correlation. 
It was confirmed in paper \cite{DI}.
One should also mention independent calculation of $P(3)$ in
\cite{BGSS}.
One can  express $P(3)$ in terms of next to the nearest neighbor
correlation

\be
<\;S^z_{i}S^z_{i+2}\;>\;=\;2\,P(3)\;-\;2\,P(2)\;+\;{1\over 2}\,P(1)
\label{G2a}
\ee

The calculation of $P(3)$ and
$P(4)$ is discussed in this paper.

{\bf The expression above for $P(4)$ is our main result here.}

We briefly announced our results in \cite{BK}, here we provide the detailed derivation.
The plan of the paper is as follows. In the next section we discuss
a general procedure of the calculation of $P(n)$. We also show how
this scheme works for $P(2)$.
In the Appendices A and B we describe in detail the calculation
of $P(3)$ and $P(4)$ respectively by means of the technique elaborated
in the Section 2. The main results are summarized in the conclusion.

\section{General discussion of the calculation of $P(n)$}

There are several different approaches to investigate $P(n)$:

\begin{itemize}

\item
{\bf Representation of correlation functions as determinants of
Fredholm integral operators} .This approach is based on following steps:

i. Quantum correlation function should be represented as a determinant of a 
Fredholm integral operators of a special type. We call these operators {\it integrable}
integral operators. 

ii. The determinant can be described by completely integrable equation of Painleve type.

iii. Asymptotic of correlation function  [and the determinant] can be described by Riemann-Hilbert problem.

This approach was discovered in \cite{iiks}, it is  described in detail in the book \cite{KIB} .

It is interesting to mention that this approach was successfully applied also to matrix models \cite{tw}

\item
{\bf Vertex operator approach} was developed in Kyoto by Foda, Jimbo, Miki, Miwa and Nakayashiki.
This approach is based on study of representations of infinite dimensional quantum group
 $U_{q}\widehat {SL(2)}$, see \cite{JMMN} .

\end{itemize}

We shall use the integral representation obtained in
\cite{KIEU} 

\be
P(n)=\int_C {d\lambda_1\over 2\pi i\lambda_1}
\int_C {d\lambda_2\over 2\pi i\lambda_2}\ldots
\int_C {d\lambda_n\over 2\pi i\lambda_n}
\prod_{a=1}^n (1+{i\over\lambda_a})^{n-a}
({\pi\lambda_a\over\sinh{\pi\lambda_a}})^n
\prod_{1\le k<j\le n}{\sinh{\pi(\lambda_j-\lambda_k)}
\over\pi(\lambda_j-\lambda_k-i)}.
\label{intPn}
\ee
The contour $C$ in each integral goes parallel to the real axis
with the imaginary part between \\
$0$ and $-i$.
In the frame of algebraic Bethe Ansatz this formula was derived in 
\cite{kmt}.
Recently such formula was  generalized  in paper
\cite{GK} to the case, where averaging is done over arbitrary Bethe state [ with no strings ]
 instead of  anti-ferromagnetic state.

Let us describe in general a strategy that may be used for the 
calculation of $P(n)$.
The integral formula (\ref{intPn}) can be easily represented
as follows:
\be
P(n)=
\prod_{j=1}^n\int_{C}{d\lambda_j\;\over 2\pi i }\;
U_n(\l_1,\ldots,\l_n)\;T_n(\l_1,\ldots,\l_n)
\label{intPna}
\ee
where
\be
U_n(\l_1,\ldots,\l_n)\;=\;\pi^{{n(n+1)\over 2}}\>
{\prod_{1\leq k < j\leq n}\sinh{\pi(\lambda_j-\lambda_k)}
\over
\prod_{j=1}^n\sinh^n{\pi\lambda_j}
}
\label{U_n}
\ee
and
\be
T_n(\l_1,\ldots,\l_n)\;=\;
{\prod_{j=1}^n\l_j^{j-1}(\lambda_j+i)^{n-j}\over
\prod_{1\leq k < j\leq n}(\lambda_j-\lambda_k-i)
}
\label{T_n}
\ee

First of all, let us note that in principle the contour $C$ can be
chosen between $0$ and $-i$ arbitrary. Let us denote $C_{\a}$ the
contour that goes from $i\a - \infty$ to $i\a +\infty$. 
In what follows
it will be convenient to choose $\a=-1/2$ i.e. to integrate over
the contour $C_{-1/2}$.

As appeared we can make a lot of simplifications without taking
integrals but using some simple observations and properties of
the function in the r.h.s. of (\ref{intPna}) which has to be 
integrated.

Let us define a "weak" equality $\sim$. Namely, let us say that
two functions $F_n(\l_1,\ldots,\l_n)$ and $G_n(\l_1,\ldots,\l_n)$ 
are "weakly" equivalent
\be
F_n(\l_1,\ldots,\l_n)\;\sim\;G_n(\l_1,\ldots,\l_n)
\label{sim}
\ee
if 
\be
\prod_{j=1}^n\int_{C_{-1/2}}{{d\l_j}\over{2\pi i}}\>
U_n(\l_1,\ldots,\l_n)\;
(F_n(\l_1,\ldots,\l_n)\;-\;G_n(\l_1,\ldots,\l_n))\>=\>0.
\ee

Let us also introduce a "canonical" form of the function by the
following formula
\be
T_n^c(\l_1,\ldots,\l_n)\;=\;
\sum_{j=0}^{[{n\over 2}]}\;
P_j^{(n)}
\prod_{k=1}^j{1\over{\l_{2k}-\l_{2k-1}}}
\label{T_n^c}
\ee
where $P_j^{(n)}$ are some polynomials of the form
$$
P_j^{(n)}\equiv 
P_j^{(n)}(\l_1,\l_3,\ldots,\l_{2j-1}|\l_{2j+1},\l_{2j+2},\ldots,\l_n)
\;=\;
$$
\be
=\;\sum_
{\tiny {\begin{array}{lcr}
\phantom{a} & 0 \le i_1, i_3, \ldots, i_{2j-1} 
\le n-2 & \phantom{a}\\
\phantom{a} & 0 \le i_{2j+1}< i_{2j+2}< \ldots < i_n\le n-1 &\phantom{a}
\end{array}
}}
C_{i_1,\quad i_3,\;\ldots,\; i_{2j-1}}^
{ i_{2j+1}, i_{2j+2},\ldots, i_n}\quad
\l_1^{i_1}\>\l_3^{i_3}\>\ldots\>\l_{2j-1}^{i_{2j-1}}\>
\l_{2j+1}^{i_{2j+1}}\>\l_{2j+2}^{i_{2j+2}}\>\ldots\> \l_n^{i_n}
\label{P_j}
\ee
where 
$$
C_{i_1,\quad i_3,\;\ldots,\; i_{2j-1}}^
{ i_{2j+1}, i_{2j+2},\ldots, i_n}=
i^{\b}\hat C_{i_1,\quad i_3,\;\ldots,\; i_{2j-1}}^
{ i_{2j+1}, i_{2j+2},\ldots, i_n}
$$
with $\b=0$ or $1$ in accordance with the equality
$$
\b+i_1+i_3+\ldots+i_{2j-1}+i_{2j+1}+i_{2j+2}+\ldots+i_n
\equiv j+n\quad\mbox{mod}\quad 2
$$
and some rational numbers $\hat C_{i_1,\quad i_3,\;\ldots,\; i_{2j-1}}^
{ i_{2j+1}, i_{2j+2},\ldots, i_n}$.

This form has some arbitrariness
because if we substitute $\l_j=x_j-i/2$ where all $x_j$ are real
then it is easy to see
that the function $\tilde U_n(x_1,\ldots,x_n)=
U_n(x_1-i/2,\ldots,x_n-i/2)$ transforms when 
$\{x_1,\ldots,x_n\}\rightarrow\{-x_1,\ldots,-x_n\}$
as follows
\be
\tilde U_n(-x_1,\ldots,-x_n)\;=\;(-1)^{n(n-1)\over 2}\>\tilde U_n(x_1,\ldots,x_n).
\label{tildeU}
\ee
Therefore any function $\tilde F_n(x_1,\ldots,x_n)$ that satisfies
\be
\tilde F_n(-x_1,\ldots,-x_n)=(-1)^{{n(n-1)\over 2}+1}\;\tilde F_n(x_1,\ldots,x_n)
\label{tildeF}
\ee
being integrated makes zero contribution
$$
\prod_{j=1}^n\int_{-\infty}^{\infty}{{dx_j}\over{2\pi i}}
\tilde U_n(x_1,\ldots,x_n)\;\tilde F_n(x_1,\ldots,x_n)\;=\;0.
$$
It means that in order to get a nonzero result one should have the
function $\tilde F_n(x_1,\ldots,x_n)$ of the same parity as 
of the function $\tilde U_n(x_1,\ldots,x_n)$. 
Then if we re-expand the form (\ref{T_n^c}) in terms of variables $x_j$
instead of $\l_j$ we can fix the arbitrariness by imposing some 
additional constraints, namely,
$$
\tilde P_j^{(n)}(x_1,x_3,\ldots,x_{2j-1}|x_{2j+1},x_{2j+2},\ldots,x_n)
\;=\;
$$
$$
=\;P_j^{(n)}(x_1-i/2,x_3-i/2,\ldots,x_{2j-1}-i/2|
x_{2j+1}-i/2,x_{2j+2}-i/2,\ldots,x_n-i/2)\;=\;
$$
\be
\sum_
{\tiny {\begin{array}{lcr}
\phantom{a} & 0 \le i_1, i_3, \ldots, i_{2j-1} 
\le n-2 & \phantom{a}\\
\phantom{a} & 0 \le i_{2j+1}< i_{2j+2}< \ldots < i_n\le n-1 &
\phantom{a}\\
\phantom{a} & i_1+i_3+\ldots+i_{2j-2}+i_{2j+1}+i_{2j+2}+\ldots+i_n
\equiv j+n \>\mbox{mod}\>2
\end{array}
}}
\tilde C_{i_1,\quad i_3,\;\ldots,\; i_{2j-1}}^
{ i_{2j+1}, i_{2j+2},\ldots, i_n}
\quad
x_1^{i_1}\>x_3^{i_3}\>\ldots\>x_{2j-1}^{i_{2j-1}}\>
x_{2j+1}^{i_{2j+1}}\>x_{2j+2}^{i_{2j+2}}\>\ldots\> x_n^{i_n}
\label{P_ja}.
\ee
In comparison with the coefficients 
$C_{i_1,\quad i_3,\;\ldots,\; i_{2j-1}}^
{ i_{2j+1}, i_{2j+2},\ldots, i_n}$ which can be pure imaginary 
all the coefficients $\tilde C_{i_1,\quad i_3,\;\ldots,\; i_{2j-1}}^
{ i_{2j+1}, i_{2j+2},\ldots, i_n}$ are real and rational numbers.

So we can expect that the function 
\be
\tilde T_n^c(x_1,\ldots,x_n)\equiv T_n^c(x_1-i/2,\ldots,x_n-i/2)
\label{tildeT_n^c}
\ee
should satisfy the following property
\be
\tilde T_n^c(-x_1,\ldots,-x_n)=(-1)^{n(n-1)\over 2}\;\tilde T_n^c(x_1,\ldots,x_n)
\label{tildeT_n^ca}
\ee
Below the property (\ref{tildeT_n^ca}) will be implied when we will 
speak about the "canonical" form (\ref{T_n^c}-\ref{P_j}).
Besides, one can note that the function $\tilde T_n^c(x_1,\ldots,x_n)$
should be real for real variables $x_j$.

Our hypothesis is that for any $n$ one can reduce the function $T_n$
defined by (\ref{T_n}) to the canonical form i.e.
there exist polynomials $P_j$ in (\ref{T_n^c})
such that 
\be
T_n(\l_1,\ldots,\l_n)\;\sim\;T_n^c(\l_1,\ldots,\l_n).
\label{T_nT_n^c}
\ee
Unfortunately, for the moment we do not have a proof of this 
statement for any $n$ 
but we will demonstrate below how it works for $n=2,3,4$.

In fact, the problem of the calculation of $P(n)$ given by the
integral (\ref{intPna}) can be reduced to the two steps. The
first step corresponds to the obtaining of the "canonical" form
for $T_n$. The second step is the calculation of the integral
by means of this "canonical" form.

To do this one needs the following simple facts:

I. $\quad$ 
Since the function $U_n(\l_1,\ldots,\l_n)$ 
is antisymmetric with respect to transposition of any pair of
integration variables, say, $\l_j$ and $\l_k$ the following
integral
\be
\prod_{j=1}^n\int_{C}{d\lambda_j\;\over 2\pi i }\;
U(\l_1,\ldots,\l_n)\;S(\l_1,\ldots,\l_n)\;=\;0
\label{int}
\ee
if the function $S$ is symmetric for at least one pair of $\l$-s.
Therefore for an arbitrary function $F_n(\l_1,\ldots,\l_n)$ 
one can transpose any pair of $\l$-s taking into consideration
appearance of additional sign because of the antisymmetry of
$U_n(\l_1,\ldots,\l_n)$. For example, if one transposes $\l_j$
with $\l_k$ one gets
\be
F_n(\ldots,\l_j,\ldots,\l_k,\ldots)\;\sim\;
-F_n(\ldots,\l_k,\ldots,\l_j,\ldots).
\label{transp}
\ee

II. $\quad$ The reduction of the power of denominator for $T_n$
is based on two relations which can be checked directly
$$
{1\over \l_k-\l_l-i}\;{1\over \l_j-\l_l-i}\;{1\over \l_j-\l_k-i}\;=
$$
\be
i\;{1\over \l_j-\l_l-i}\;{1\over \l_j-\l_k-i}\;+\;
i\;{1\over \l_k-\l_l-i}\;{1\over \l_j-\l_l-i}\;-\;
i\;{1\over \l_k-\l_l-i}\;{1\over \l_j-\l_k-i}
\label{re1}
\ee
\be
\prod_{k=1}^{j-1}{1\over \l_j-\l_k-i}=
\sum_{k=1}^{j-1}{1\over \l_j-\l_k-i}\prod_{
\tiny
\begin{array}{lcr}
\phantom{a} & l=1 &\phantom{a}\\
\phantom{a} & l\ne k &\phantom{a}
\end{array}
}^{j-1}
{1\over \l_k-\l_l}
\label{re2}.
\ee
In Appendices A and B we will show how the reduction can be performed
for $n=3$ and $n=4$. Unfortunately,
so far we have not succeeded in finding a result for general $n$.

III. $\quad$ The ratio 
\be
{{T_{n+1}(\l_1,\ldots,\l_{n+1})}\over
{T_{n}(\l_1,\ldots,\l_{n})}}\;=\;
{\prod_{j=1}^n(\l_j+i)\>\l_{n+1}^n\over
\prod_{j=1}^n(\l_{n+1}-\l_j-i)}
\label{ratio}
\ee
is symmetric with respect to any permutation of $\l_1,\ldots,\l_n$. 
Therefore
the relation (\ref{ratio}) allows us to use the result $T_n$ also for
derivation of $T_{n+1}$ if this result was obtained by applying 
the relations (\ref{transp}-\ref{re2}) from I and II.

IV. {\bf Proposition 1}\\
Let the function $f(\l_1,\ldots,\l_n)$ have only poles of the form
$1/(\l_j-\l_k+i a)$ with $a$ an integer i.e. the product
$U_n(\l_1,\ldots,\l_n)f(\l_1,\ldots,\l_n)$ does not have poles of
that kind. Then 
\be
\l_j^m\,f(\ldots,\l_j,\ldots)\;\sim\;-(\l_j+i)^m\,
f(\ldots,\l_j+i,\ldots)
\label{l^m}
\ee
where $m$ is an integer and $m\ge n$.

$\quad$\\
{\it Proof}\\
Let us suppose that all variables $\l_1,\ldots,\l_{j-1},\l_{j+1},\ldots,
\l_n$ are fixed. Extracting from $U_n(\l_1,\ldots,\l_n)$ 
the function which depends on $\l_j$ one gets
$$
\int_{C_{-1/2}}{d\l_j\over 2\pi i}
{\prod_{k\ne j}\sinh{\pi(\l_j-\l_k)}\over 
\sinh^n{\pi\l_j}}
\l_j^m\;w(\l_j)\;=\;
-\int_{C_{-3/2}}{d\l_j\over 2\pi i}
{\prod_{k\ne j}\sinh{\pi(\l_j-\l_k)}\over 
\sinh^n{\pi\l_j}}
(\l_j+i)^m\;w(\l_j+i)\;=
$$
$$
=\;-(\int_{C_{-3/2}}-\int_{C_{-1/2}}+\int_{C_{-1/2}})
{d\l_j\over 2\pi i}
{\prod_{k\ne j}\sinh{\pi(\l_j-\l_k)}\over 
\sinh^n{\pi\l_j}}
(\l_j+i)^m\;w(\l_j+i)
$$
where $w(\l_j)=f(\ldots,\l_j,\ldots)$. The first step here was
to shift integration variable $\l_j\rightarrow\l_j+i$ and to use
the fact that $\sinh{\pi(x+i)}=-\sinh{\pi x}$. The two first integrals
in the last expression are equal to a contour integral around 
the point $\l_j=-i$ in a complex plane of the variable $\l_j$. 
Since, $m\ge n$ the term $(\l_j+i)^m$ which is in the numerator 
and corresponds to a zero of order $m$ compensates the pole
from the term $\sinh^n{\pi\l_j}$ in the denominator. Therefore the
contribution of those two integrals is zero and we immediately come to 
the statement (\ref{l^m}).

\vspace{0.3cm}

One can get two useful corollaries from the proposition 1.

$\quad$\\
{\it Corollary 1}
\be
\l_j^m\;g(\l_1,\ldots,\hat\l_j,\ldots,\l_n)\;\sim\;
{(-i)\over 2}\sum_{k=0}^{m-1}\l_j^k(\l_j+i)^{m-1-k}\;
g(\l_1,\ldots,\hat\l_j,\ldots,\l_n)
\label{cor1}
\ee
where the function $g(\l_1,\ldots,\hat\l_j,\ldots,\l_n)$ does not
depend on $\l_j$ and as above it is implied that $m\ge n$.

$\quad$\\
{\it Proof}\\
The relation (\ref{cor1}) is easy to derive using the relation
$(\l_j^m+(\l_j+i)^m)g(\l_1,\ldots,\hat\l_j,\ldots,\l_n)\sim 0$
or equivalently 
$\l_j^m g(\l_1,\ldots,\hat\l_j,\ldots,\l_n)\sim 1/2(\l_j^m-(\l_j+i)^m)
g(\l_1,\ldots,\hat\l_j,\ldots,\l_n)$.

$\quad$\\
{\it Corollary 2}
$$
{\l_j^{m-1}\over \l_k-\l_j}\;
g(\l_1,\ldots,\hat\l_k,\ldots,\hat\l_j,\ldots,\l_n)\;\sim\;
$$
\be
\sim\;{i\over m}\Bigl(\sum_{l=2}^m 
{\Bigl(
\begin{array}{c}
m\\
l
\end{array}
\Bigr)\;
i^l\l_j^{m-l}\over
\l_k-\l_j}
\>+\>
\sum_{l=0}^{m-1}\l_k^l(\l_j+i)^{m-1-l}\Bigr)
g(\l_1,\ldots,\hat\l_k,\ldots,\hat\l_j,\ldots,\l_n)
\label{cor2}
\ee
where 
$$
\Bigl(
\begin{array}{c}
m\\
l
\end{array}
\Bigr)={m!\over l!\,(m-l)!}
$$
is binomial coefficient and the function 
$g(\l_1,\ldots,\hat\l_k,\ldots,\hat\l_j,\ldots,\l_n)$ does not depend
on $\l_k$ and $\l_j$ and $m\ge n$.

$\quad$\\
{\it Proof}\\
Using the proposition 1 we get
$$
{\l_j^{m}\over \l_k-\l_j}\>
g(\l_1,\ldots,\hat\l_k,\ldots,\hat\l_j,\ldots,\l_n)\;\sim\;
-{(\l_j+i)^{m}\over \l_k-\l_j-i}\>
g(\l_1,\ldots,\hat\l_k,\ldots,\hat\l_j,\ldots,\l_n)\;=\;
$$
$$
=\;\Bigl(-{\l_k^{m}\over \l_k-\l_j-i}+
{\l_k^{m}-(\l_j+i)^{m}\over \l_k-\l_j-i}\Bigr)\>
g(\l_1,\ldots,\hat\l_k,\ldots,\hat\l_j,\ldots,\l_n)\;\sim\;
$$
$$
\sim\;\Bigl({(\l_k+i)^{m}\over \l_k-\l_j}+
{\l_k^{m}-(\l_j+i)^{m}\over \l_k-\l_j-i}\Bigr)\>
g(\l_1,\ldots,\hat\l_k,\ldots,\hat\l_j,\ldots,\l_n)\;\sim\;
$$
$$
\sim\;\Bigl({(\l_j+i)^{m}\over \l_k-\l_j}+
{\l_k^{m}-(\l_j+i)^{m}\over \l_k-\l_j-i}\Bigr)\>
g(\l_1,\ldots,\hat\l_k,\ldots,\hat\l_j,\ldots,\l_n)\;\sim\;
$$
or 
$$
\Bigl({(\l_j+i)^{m}-\l_j^m\over \l_k-\l_j}+
{\l_k^{m}-(\l_j+i)^{m}\over \l_k-\l_j-i}\Bigr)\>
g(\l_1,\ldots,\hat\l_k,\ldots,\hat\l_j,\ldots,\l_n)\;\sim\;0.
$$
Then expanding both numerators according to the formulae
$$
(\l_j+i)^{m}-\l_j^m=\sum_{l=1}^m
\Bigl(
\begin{array}{c}
m\\
l
\end{array}
\Bigr)\;i^l\l_j^{m-l}
$$
$$
\l_k^m-(\l_j+i)^m=(\l_k-\l_j-i)\sum_{l=0}^{m-1}\l_k^l(\l_j+i)^{m-1-l}
$$
we arrive at the formula (\ref{cor2}).

With the help of the corollaries 1 and 2 one can effectively 
reduce the power of the numerator in $T_n$.


V. $\quad$ For the calculation of integrals we need the following\\

{\bf Proposition 2} $\quad$ Let the integral
$\int_{C_{-1/2}}{d\l\over 2\pi i}\>e^{i\d\l}\>{f(\l+i\, N)\over \sinh^n{\pi\l}}$
be convergent for any real number $\d\ge 0$ and an integer $N$.
Further, let the function $f(\l)$ be analytic in the whole complex plane $\l$
and satisfy the following two conditions
\be
\lim_{R\rightarrow\infty} |e^{-n\,\pi\, R}\; f(i y\,-\,i/2\,\pm\, R)|\,=\,0,
\label{limR}
\ee
\be
\lim_{N\rightarrow\infty} |e^{-\d'\,N}\; {f(x\,-\,i/2\,\pm\,i N)\over \cosh^n{\pi x}}|\,=\,0
\label{limN}
\ee
where the first limit is uniform in $y$ , when  $y\in [0,N]$. The second limit
is uniform in $x$ for  any real  $x$.
The value $\d'$ is a fixed real positive number ($\d'>0$). Then
\beq
&{\ds \int_{C_{-1/2}}{d\l\over 2\pi i}\>{f(\l)\over \sinh^n{\pi\l}}\;=\;
\lim_{\d\rightarrow 0^+}{d^{(n)}(\e)}_{\e\rightarrow 0}
\sum_{l=0}^{\infty}(-1)^{ln}\;e^{-\d l}\;f(il+\e)\;=}&
\label{int1}\\
&{\ds =\; -\lim_{\d\rightarrow 0^+}{d^{(n)}(\e)}_{\e\rightarrow 0}
\sum_{l=1}^{\infty}(-1)^{ln}\;e^{-\d l}\;f(-il+\e)}&
\label{int2}
\eeq
where a differential operator $d^{(n)}(\e)$ looks as follows
\be
d^{(n)}(\e)\;=\;{1\over \pi^n(n-1)!}
\sum_{l=0}^{n-1}
\Bigl(
\begin{array}{c}
n-1\\
l
\end{array}
\Bigr)
{\Bigl({\partial\over\partial\a}\Bigr)}^{n-1-l}_{\a\rightarrow 0}
{\Bigl(\sum_{0\le 2k<n}{{(\pi\a)}^{2k}\over (2k+1)!}\Bigr)}^{-n}
{\Bigl({\partial\over\partial\e}\Bigr)}^{l}.
\label{dn}
\ee
In particular, for $n=2, 3, 4$ 
\be
d^{(2)}(\e)\;=\;{1\over\pi^2}{\partial\over\partial\e}
\label{d2}
\ee
\be
d^{(3)}(\e)\;=\;-{1\over 2\pi}\;{(1-{1\over\pi^2}
{\partial^2\over{{\partial\e}^2}})}
\label{d3}
\ee
\be
d^{(4)}(\e)\;=\;-{2\over 3\pi^2}\;
{({\partial\over{{\partial\e}}}
-{1\over 4\pi^2}{\partial^3\over{{\partial\e}^3}})}.
\label{d4}
\ee


$\quad$\\
{\it Proof} $\quad$ Let 
\be
F(\l,N)\,=\,\int_{C_N}{d\l\over 2\pi i}\,e^{i\d\l}{f(\l)\over \sinh^n{\pi\l}}
\label{FlN}
\ee
where $\d>0$ and $C_N$ is a rectangular contour shown in Fig. 1.

\begin{picture}(500,200)
\put(150,100){
\begin{picture}(450,150)
\thicklines
\put(-10,0){\vector(1,0){220}}
\put(100,-30){\vector(0,1){120}}
\multiput(10,-10)(0,0.5){2}{\line(1,0){180}}
\multiput(110,-10)(0,1){2}{\vector(1,0){20}}
\multiput(10,70)(0,0.5){2}{\line(1,0){180}}
\multiput(80,70)(0,1){2}{\vector(-1,0){20}}
\multiput(10,-10)(1,0){2}{\line(0,1){80}}
\multiput(10,45)(1,0){2}{\vector(0,-1){20}}
\multiput(190,-10)(1,0){2}{\line(0,1){80}}
\multiput(190,20)(1,0){2}{\vector(0,1){20}}
\multiput(100,-20)(0,20){3}{\circle*{5}}
\multiput(100,60)(0,20){2}{\circle*{5}}
\multiput(105,30)(0,4){4}{\circle*{1}}
\put(50,-20){$C_{-1/2}$}
\put(50,80){$-C_{N-1/2}$}
\put(198,20){$C^+_{\infty}$}
\put(-15,20){$C^-_{\infty}$}
\put(105,5){\tiny $0$}
\put(105,20){\tiny $i$}
\put(105,60){\tiny $(N-1)i$}
\put(35,-60){Fig. 1 $\quad$ The contour $C_N$}
\end{picture}
}
\end{picture}

The contours $C_{-1/2}$ and $-C_{N-1/2}$ correspond to lower and upper horizontal parts of
the contour $C_N$ respectively (sign $-$ is because the contour $C_{N-1/2}$ should
be taken in the opposite direction). The contours $C^{+}_{\infty}$ and
$C^{-}_{\infty}$  correspond to
the right and left vertical parts of the contour $C_N$ and have real parts $+\infty$
and $-\infty$ respectively.
Due to the Cauchy theorem one has
\be
F(\d,N)\;=\;F_{-1/2}\;-\;F_{N-1/2}\;+\;F_+\;+\;F_-\;=\;
{d^{(n)}(\e)}_{\e\rightarrow 0}\;\sum_{l=0}^{N-1}(-1)^{ln}\,e^{-\d l+i\d\e}\;
f(il\,+\,\e)
\label{FdN}
\ee
where 
\beq
&{\ds F_{-1/2}\;=\;
\int_{C_{-1/2}}{d\l\over 2\pi i}\>e^{i\d\l}\>{f(\l)\over \sinh^n{\pi\l}}}&
\label{F1/2}\\
&{\ds F_{N-1/2}\;=\;
\int_{C_{N-1/2}}{d\l\over 2\pi i}\>e^{i\d\l}\>{f(\l)\over \sinh^n{\pi\l}}}&
\label{FN/2}\\
&{\ds F_{\pm}\;=\;
\int_{C^{\pm}_{\infty}}{d\l\over 2\pi i}\>e^{i\d\l}\>{f(\l)\over \sinh^n{\pi\l}}}&
\label{Fpm}
\eeq
The r.h.s. of the formula (\ref{FdN}) is a result of
the calculation of residues corresponding to zeros of the denominator 
$\sinh^n{\pi\l}$ which are placed inside the contour $C_N$.


The first step is to prove that the integrals over the contours $C^{\pm}_{\infty}$
\be
F_{\pm}=0.
\label{Fpm0}
\ee
Actually one has
$$
F_{\pm}\;=\;\lim_{R\rightarrow\infty}
\int_{C^{\pm}_{R}}{d\l\over 2\pi i}\>e^{i\d\l}\>{f(\l)\over \sinh^n{\pi\l}}\;=\;
$$
\be
=\;{1\over 2\pi i}
\lim_{R\rightarrow\infty}\pm\int_0^N {dy\over (-i)^n\cosh^n{\pi(\pm R\,+\,iy)}}
e^{i\d(\pm R\,+\,iy\,-\,i/2)}f(\pm R\,+\,iy\,-\,i/2)
\label{Fpm1}
\ee
where vertical contours $C^{\pm}_R$ are defined
as follows: $\{\l=\pm R\,+\,iy\,-\,i/2;\quad y\in [0,N]\}$.
Then
$$
|F_{\pm}|\;\le\;
{1\over 2\pi}\lim_{R\rightarrow\infty}
\int_0^N {dy\over \sinh^n{\pi R}}
e^{-\d y} |f(\pm R\,+\,iy\,-\,i/2)|\;=
$$
\be
=\;{2^{n-1}\over \pi}\lim_{R\rightarrow\infty}
\int_0^N {dy e^{-\d y}\over (1\,-\,e^{-2\pi R})^n}
e^{-n\pi R} |f(\pm R\,+\,iy\,-\,i/2)|
\label{Fpm2}
\ee
where we have used a simple fact that for $R>0$
$$
|\cosh{\pi (\pm R\,+\,iy)}|\ge \sinh{\pi R}.
$$
The uniform character of the limit (\ref{limR}) allows to interchange
the order of the integration over $y$ and the limiting procedure $R\rightarrow\infty$.
Indeed, the condition (\ref{limR}) means that for any small real number $\e$ there
exists a real $R_{\e}$ which is independent of $y$ such that for any
$R>R_{\e}$
$$
e^{-n\pi R} |f(\pm R\,+\,iy\,-\,i/2)|\;<\;\e.
$$
Therefore for $R>R_{\e}$ 
$$
\int_0^N {dy e^{-\d y}\over (1\,-\,e^{-2\pi R})^n}
e^{-n\pi R} |f(\pm R\,+\,iy\,-\,i/2)|\;<\;{\e\over (1-e^{-2\pi R})^n}
\int_0^N dy e^{-\d y}\;=
$$
$$
=\;{\e\over (1-e^{-2\pi R})^n}{1\,-\,e^{-N\d}\over\d}\;<
\;{\e\over (1-e^{-2\pi R_{\e}})^n}{1\,-\,e^{-N\d}\over\d}
$$
Hence we have got that
$$
\lim_{R\rightarrow\infty}\int_0^N {dy e^{-\d y}\over (1\,-\,e^{-2\pi R})^n}
e^{-n\pi R} |f(\pm R\,+\,iy\,-\,i/2)|\;=\;0
$$
and we come to the statement (\ref{Fpm0}). 

The next our step is to prove that for a fixed real $\d>0$
\be
\lim_{N\rightarrow\infty}F_{N-1/2}\;=\;0
\label{FN0}
\ee
Indeed,
$$
|F_{N-1/2}|\;=\;
|\int_{C_{N-1/2}}{d\l\over 2\pi i}\>e^{i\d\l}\>{f(\l)\over \sinh^n{\pi\l}}|\;=
|(-1)^{Nn}
\int_{C_{-1/2}}{d\l\over 2\pi i}\>e^{i\d(\l\,+\,iN)}\>{f(\l\,+\,iN)\over \sinh^n{\pi\l}}|
\;=
$$
\be
=\;{e^{-\d N}\over 2\pi}|\int_{C_{-1/2}}{d\l\over \sinh^n{\pi\l}}
\>e^{i\d\l}\> f(\l\,+\,iN)|\;\le
\;{e^{-\d N}\over 2\pi}\int_{-\infty}^{\infty}{dx\over \cosh^n{\pi x}}
|f(x\,-\,i/2\,+\,i N)|
\label{FN1}
\ee
As above due to the uniform character of the limit (\ref{limN}) one can interchange 
the integration over $x$ and the limit $N\rightarrow\infty$. Therefore the last
expression in (\ref{FN1}) tends to zero when $N\rightarrow\infty$ and we come to
the statement (\ref{FN0}).

So for a fixed $\d > 0$ we conclude from (\ref{FdN}) that
$$
\lim_{N\rightarrow\infty} F(\d,N)\;=\;\int_{C_{-1/2}}{d\l\over 2\pi i}
e^{i\d\l}{f(\l)\over \sinh^n{\pi\l}}\;=\;
{d^{(n)}(\e)}_{\e\rightarrow 0}\;\sum_{l=0}^{\infty}(-1)^{ln}\,e^{-\d l+i\d\e}\;
f(il\,+\,\e).
$$
The convergence of the sum in r.h.s. of the last expression is guaranteed
by the convergence of the integral in the l.h.s.
Finally, after taking the limit $\lim_{\d\rightarrow 0^+}$ we come to
the formula (\ref{int1}). Let us note that generally speaking we can
not interchange the order of the limits $N\rightarrow\infty$ and
$\d\rightarrow 0^+$.
 
Another form (\ref{int2}) can be proved 
in a similar way if considering a contour which is analogous to $C_N$ but 
placed in a lower half-plane and a real number $\d$ should be taken
negative. It completes the proof of the proposition 2.


\vspace{0.2cm}
$\quad$\\
Using the proposition 2 we can get the following \\
{\bf Proposition 3}
\be
\prod_{j=1}^n\int_{C_{-1/2}}{d\l_j\;\over 2\pi i }
U_n(\l_1,\ldots,\l_n)\>F_n(\l_1,\ldots,\l_n)
\;=\;D^{(n)}\;\tilde F_n(\e_1,\ldots,\e_n)
\label{IS}
\ee
where the multiple integral is convergent and 
the product $U_n(\l_1,\ldots,\l_n)\>F_n(\l_1,\ldots,\l_n)$
does not have any other poles besides the poles of the denominator
$\prod_{j=1}^n\sinh^n{\pi \l_j}$ of the function
$U_n(\l_1,\ldots,\l_n)$. The function 
\be
G^{(j)}_n(\l_1,\ldots,\l_n)\;=\;
{\prod_{1\leq k < l\leq n}\sinh{\pi(\l_l-\l_k)}\over
\prod_{k\neq j}\sinh^n{\pi \l_k}}
F_n(\l_1,\ldots,\l_n)
\label{Gnj}
\ee
should satisfy conditions which generalize (\ref{limR}) and (\ref{limN}) 
\be
\lim_{R\rightarrow\infty} |e^{-n\,\pi\, R}\; 
G^{(j)}_n(\l_1,\ldots,\l_{j-1},i x_j\,-\,i/2\,\pm\, R,\l_{j+1},\ldots,\l_n)|\,=\,0,\quad
j=1,\ldots,n
\label{limR1}
\ee
\be
\lim_{N\rightarrow\infty} |e^{-\d\,N}\; 
G^{(j)}_n(\l_1,\ldots,\l_{j-1},x_j\,-\,i/2\,\pm\, i N,\l_{j+1},\ldots,\l_n)|/
{\cosh^n{\pi x_j}}
\,=\,0,\quad j=1,\ldots,n
\label{limN1}
\ee
where for each $j$ both limits are uniform on real numbers 
$x_1,\ldots,x_n$ with $\l_k=x_k+i m_k/2,\quad k\neq j$  and some integers $m_k$.
For the first limit $x_j\in [0,N]$ while for the second one $x_j$ is any real number.
$D^{(n)}$ is a differential operator
\be
D^{(n)}=\pi^{n(n+1)\over 2}
\prod_{j=1}^n{d^{(n)}(\e_j)}_{\e_j\rightarrow 0}
\prod_{1\le k<j\le n}\sinh{\pi (\e_j-\e_k)}
\label{D}
\ee
and
\be
\tilde F_n(\e_1,\ldots,\e_n)\;=\;\lim_{\d_1\rightarrow 0^+}
\sum_{l_1=0}^{\infty}(-1)^{l_1}e^{-\d_1 l_1}
\ldots
\lim_{\d_n\rightarrow 0^+}\sum_{l_n=0}^{\infty}(-1)^{l_n}e^{-\d_n l_n}
F_n(i\,l_1\,+\,\e_1,\ldots,i\,l_n\,+\,\e_n)
\label{F}
\ee
where each sum $\sum_{l_j=0}^{\infty}(-1)^{l_j}e^{-\d_j l_j}
F_n(\l_1,\ldots,\l_{j-1},i\,l_j\,+\,\e_j,\l_{j+1},\ldots,\l_n)$ should be convergent
uniformly in other arguments $\l_1,\ldots,\l_{j-1},\l_{j+1},\ldots,\l_n$
($\l_k=x_k+i m_k/2,\quad k\neq j$).


{\it Proof} $\quad$ The formula (\ref{IS}) can be got by the recursive
application of the formula (\ref{int1}) to each integral in the l.h.s. of (\ref{IS})
and by taking into account the manifest form (\ref{U_n}) of the
function $U_n(\l_1,\ldots,\l_n)$. After $j-1$-th application one can interchange
the integration over $\l_{j}$ with the sums over $l_1,\ldots,l_{j-1}$ due to
the uniform convergence of these sums. Due to the conditions (\ref{limR1}) and
(\ref{limN1}) one can apply the proposition 2 when integrating over the variable $\l_j$.

Let us note that each sum $\sum_{l_j=0}^{\infty}(-1)^{l_j}e^{-\d_j l_j}
F_n(\ldots,i\l_j+\e_j,\ldots)$ in the formula (\ref{F}) can be 
substituted by 
$-\sum_{l_j=1}^{\infty}(-1)^{l_j}e^{-\d_j l_j}F_n(\ldots,-i\l_j+\e_j,\ldots)$
corresponding to the choice of the contour in the lower half-plane.
In what follows we will use this fact depending on a convenience.

\vspace{0.5cm}

Let us consider a special class of functions
$\tilde F_n(\e_1,\ldots,\e_n)$ that does not have a singularity 
when $\e_j\rightarrow 0$ for $j=1,\ldots,n$.
In this case one can expand $\tilde F_n(\e_1,\ldots,\e_n)$ into the
infinite series on powers of $\e$-s. 
We have checked that for $n\le 4$ 
the differential operator $D^{(n)}$ given by 
(\ref{D}) when acting on some monomial $\e_1^{i_1}\ldots \e_n^{i_n}$
makes non-zero contribution only for monomial of the
form $\e_{\s(1)}^0\e_{\s(2)}^1\ldots \e_{\s(n)}^{n-1}$ where
$\s$ is some element of the permutation group $S_n$ of $n$
elements. More precisely, we can write
\be
D^{(n)}\;=\;{1\over\prod_{j=1}^{n-1}j!}\prod_{1\le k<j\le n}
{\Bigl({\partial\over\partial\e_k}-{\partial\over\partial\e_j}\Bigr)}_
{\vec\e\rightarrow 0}
\label{Da}
\ee
We proved it for $n\le 4$ and we use this formula only in this case.
We believe that this relation is valid for any $n$ but this fact
is still to be proven.

VI. Now let us discuss the integrals of a special form, namely, when the
function $F_n(\l_1,\ldots,\l_n)$ from the proposition 3 is rational 
on it's arguments and the function
$$
\prod_{1\leq k < j\leq n}\sinh{\pi(\l_j-\l_k)}F_n(\l_1,\ldots,\l_n)
$$
is analytic. It means that the function $F_n(\l_1,\ldots,\l_n)$ can have only
simple poles when $\l_j\rightarrow\l_k\,+\,m i$ with an integer $m$.
As it is seen from the definitions (\ref{T_n}), (\ref{T_n^c}) both the function
$T_n(\l_1,\ldots,\l_n)$ and the function $T_n^c(\l_1,\ldots,\l_n)$ are
of that form.

Let us show that the conditions of applicability of the propositions 2 and 3 are 
fulfilled for such a function $F_n(\l_1,\ldots,\l_n)$.
To show that the first condition (\ref{limR1}) is satisfied let us fix 
without loss of generality $j=1$ and
take into account that $0\leq x_1\leq N$. 
First let us consider the case when the rational function $F_n$ does not have
poles at all. Then if $m_2,\ldots,m_n$ have the same parity
$$
e^{-n\,\pi\, R}\; 
|G^{(1)}_n(i x_1\,-\,i/2\,\pm\, R,x_2\,+\,i m_2/2,\ldots,x_n\,+\,i m_n/2)|\;=
$$ 
$$
=\;e^{-n\,\pi\, R}\;\prod_{k=2}^n|\sinh{\pi (x_k\mp R-i x_1)}|
{\prod_{1<k<l\leq n} |\sinh{\pi (x_l-x_k)}|\over \prod_{k=2}^n \cosh^n{\pi x_k}}
$$
$$
|F_n(i x_1\,-\,i/2\,\pm\, R,x_2\,+\,i m_2/2,\ldots,x_n\,+\,i m_n/2)|\;\leq
$$
$$
\leq\;e^{-n\,\pi\, R}\;
2^{(n-2)(n-1)\over 2}\prod_{k=2}^n {|\sinh{\pi (x_k\mp R-i x_1)}|\over \cosh^2{\pi x_k}}
|F_n(i x_1\,-\,i/2\,\pm\, R,x_2\,+\,i m_2/2,\ldots,x_n\,+\,i m_n/2)|\;\leq
$$
$$
\leq\;e^{-n\,\pi\, R}\;
2^{(n-2)(n-1)\over 2}\prod_{k=2}^n{\cosh{\pi (x_k\mp R)}\over \cosh^2{\pi x_k}}
|F_n(i x_1\,-\,i/2\,\pm\, R,x_2\,+\,i m_2/2,\ldots,x_n\,+\,i m_n/2)|\;\leq
$$
$$
\leq\;e^{-n\,\pi\, R}\;2^{(n-2)(n-1)\over 2}(\cosh{\pi R}\,+\,\sinh{\pi R})^{n-1}
\prod_{k=2}^n{1\over \cosh{\pi x_k}}
$$
$$
|F_n(i x_1\,-\,i/2\,\pm\, R,x_2\,+\,i m_2/2,\ldots,x_n\,+\,i m_n/2)|\;=
$$
\be
=\;2^{(n-2)(n-1)\over 2}e^{-\pi\, R}
|F_n(i x_1\,-\,i/2\,\pm\, R,x_2\,+\,i m_2/2,\ldots,x_n\,+\,i m_n/2)|
\label{last1}
\ee
where we have used an inequality
\be
{{\ds 
\prod_{1<k<l\leq n} |\sinh{\pi (x_l-x_k)}|}\over 
{\ds \prod_{k=2}^n \cosh^n{\pi x_k} }}\;\leq\;
2^{(n-2)(n-1)\over 2}\prod_{k=2}^n {1\over \cosh^2{\pi x_k}}
\label{ineq}
\ee
which can be checked directly. For a set of arbitrary integers $m_k,\>k=2,\ldots,n$
one can repeat the derivation above also.

Since $0\leq x_1\leq N$ and $F_n$ is rational without poles one can find real
numbers $R^{\star}$ and $M(N)>0$ which is independent of $R$ such that for $R>R^{\star}$
$$
\prod_{k=2}^n{1\over \cosh{\pi x_k}}
|F_n(i x_1\,-\,i/2\,\pm\, R,x_2\,+\,i m_2/2,\ldots,x_n\,+\,i m_n/2)|\;\leq\;
R^s M(N)
$$
with some power $s$.
Hence, we get that the expression (\ref{last1}) can not exceed 
$$
2^{(n-2)(n-1)\over 2}\;e^{-\pi\, R}\;R^s M(N)
$$
which tends to zero when $R\rightarrow\infty$ independently of the 
variables $x_1,\ldots,x_n$ and we get the uniform character of 
the limit (\ref{limR1}) on these variables.
 
Let us prove the validity of the second limit (\ref{limN1}).
Indeed, for a given integer $j=1,\ldots n$ and $\l_k=x_k+i m_k/2,\quad k\neq j$ with 
integers $m_k$ of the same parity the function
$$
|e^{-\d\,N}\;
G^{(j)}_n(\l_1,\ldots,\l_{j-1},x_j\,-\,i/2\,\pm\, i N,\l_{j+1},\ldots,\l_n)|/{\cosh^n{\pi x_j}}
$$
is bounded for any real numbers $x_k\in (-\infty,\infty)$ and 
a positive integer $N$. Therefore one can use again the inequality (\ref{ineq})
in order to get
$$
\prod_{k=2}^n|\sinh{\pi (x_k- x_1 -iN)}|
{\prod_{1<k<l\leq n} |\sinh{\pi (x_l-x_k)}|\over \prod_{k=1}^n \cosh^n{\pi x_k}}\;=
\;{\prod_{1\leq k<l\leq n} |\sinh{\pi (x_l-x_k)}|\over \prod_{k=1}^n \cosh^n{\pi x_k}}\;\leq
$$
$$
\leq\; 2^{n(n-1)\over 2}{1\over \prod_{k=1}^n\cosh{\pi x_k}}
$$
Again one can repeat this for arbitrary integers $m_2,\ldots,m_n$.

Since $F_n$ is rational the maximum over the real variables $x_1,\ldots,x_n$
of the function 
$$
{1\over \prod_{k=1}^n\cosh{\pi x_k}}|F_n(x_1\,-\,i/2\,+\,iN,
x_2\,+\,i m_2/2,\ldots,x_n\,+\,i m_n/2)|
$$
can be some power of $N$, say, $N^{s'}$ multiplied with some constant which is
independent of $N$ when $N>N^{\star}$ with $N^{\star}$ is a big enough integer. 
Therefore we get the limit
$$
\lim_{N\rightarrow\infty} e^{-\d N}N^{s'}\;=\;0
$$
for any real $\d>0$ independently of variables $x_1,\ldots,x_n$
and we come to the uniform limit (\ref{limN1}).

Suppose the function $F_n(\l_1,\ldots,\l_n)$ has a
simple pole of a type $1/(\l_k-\l_l+i a_{kl})$ with an integer $a_{kl}$. Let us
restrict ourself only with the case $a_{kl}=0$ because only such poles can appear
in the expression for a canonical form (\ref{T_n^c}). 
Then we can write
$$
|{\sinh{\pi(\l_k-\l_l)}\over \l_k-\l_l}|\;\leq\;
\cases{
{\ds \pi \sinh{1}  } & if ${\ds |\l_k-\l_l| \leq {1\over \pi}  }$ \cr
{\ds \pi |\sinh{\pi(\l_k-\l_l)}| } & if ${\ds |\l_k-\l_l|> {1\over \pi} }$ \cr
}
$$
and use again the technique described above. 

Let us comment on a question of uniform convergence of a sum
$$
\sum_{l_j=0}^{\infty}(-1)^{l_j}e^{-\d_j l_j}
F_n(\l_1,\ldots,\l_{j-1},i\,l_j\,+\,\e_j,\l_{j+1},\ldots,\l_n)
$$ 
as well as how to proceed further by considering two typical examples.
Let us take for simplicity $n=2$.
The integral we need looks as follows
\be
J_2\;=\;\int_{C_{-1/2}}{d\l_2\over 2\pi i}\int_{C_{-1/2}}{d\l_1\over 2\pi i}
{\sinh{\pi (\l_2-\l_1)}\over \sinh^2{\pi\l_1}\sinh^2{\pi\l_2}}F_2(\l_1,\l_2)
\label{exn2}
\ee

{\it i} $\quad$ As a first example let us consider again the case when 
$F_2$ does not have poles at all i.e. $F_2(\l_1,\l_2)$ is some
polynomial on $\l_1$ and $\l_2$.
First let us integrate over $\l_1$ using the formula 
(\ref{int1})
\be
J_2\;=\;\int_{C_{-1/2}}{d\l_2\over 2\pi i\,\sinh^2{\pi\l_2}}
\lim_{\d_1\rightarrow 0^+} {d^{(2)}(\e_1)}_{\e_1\rightarrow 0}
\sinh{\pi (\l_2-\e_1)}\sum_{l_1=0}^{\infty} (-1)^{l_1}e^{-\d_1 l_1}F_2(il_1+\e_1,\l_2)
\label{exn21}
\ee
Since $F_2(\l_1,\l_2)$ is a polynomial then $F_2(il_1+\e_1,\l_2)$ 
is a polynomial on $l_1$ as well. Let us pick out some monomial on $l_1$ from it, say,
\be
l_1^{a}F_1(\l_2)
\label{mon1}
\ee
where $a$ is a non-negative integer and $F_1(\l_2)$ is a polynomial on $\l_2$.
Actually it has a factorized form. Therefore in this case we do not have any problems 
with the uniform convergence and the corresponding contribution into $J_2$ is as follows
$$
\int_{C_{-1/2}}{d\l_2\over 2\pi i\,\sinh^2{\pi\l_2}} F_1(\l_2)
\lim_{\d_1\rightarrow 0^+} {d^{(2)}(\e_1)}_{\e_1\rightarrow 0}
\sinh{\pi (\l_2-\e_1)}\sum_{l_1=0}^{\infty} (-1)^{l_1}e^{-\d_1 l_1}l_1^a\;=\;
$$
\be
=\;\int_{C_{-1/2}}{d\l_2\over 2\pi i\,\sinh^2{\pi\l_2}} F_1(\l_2)
{d^{(2)}(\e_1)}_{\e_1\rightarrow 0}\sinh{\pi (\l_2-\e_1)}\r(a)
\label{exn22}
\ee
where
\be
\r(a)\;=\;\lim_{\d\rightarrow 0^+}\sum_{l_1=0}^{\infty}
(-e^{-\d})^{l_1}l_1^a
\label{ro}
\ee
Let us adduce a number of first values of $\r(a)$
\be
\r(0)={1\over 2},\quad \r(1)=-{1\over 4},\quad 
\r(2)=0,\quad \r(3)={1\over 8},\quad 
\r(4)=0,\quad \r(5)=-{1\over 4}
\label{ro5}
\ee
Since $F_1(\l_2)$ is a polynomial we can treat the integral in (\ref{exn22}) in a
similar way as the integral over $\l_1$. In the very end we should calculate the
limits $\e_1\rightarrow 0$ and $\e_2\rightarrow 0$ and get the final answer for $J_2$.

{\it ii} $\quad$ The second example corresponds to an existing of a simple pole, namely,
when
\be
F_2(\l_1,\l_2)\;=\;{Q(\l_1,\l_2)\over \l_2-\l_1-i a_{12}}
\label{sec}
\ee
where $Q(\l_1,\l_2)$ is a polynomial on $\l_1,\l_2$. As above we shall consider only the case
$a_{12}=0$.
So in this case doing the first integration one gets
\be
J_2\;=\;\int_{C_{-1/2}}{d\l_2\over 2\pi i\,\sinh^2{\pi\l_2}}
\lim_{\d_1\rightarrow 0^+} {d^{(2)}(\e_1)}_{\e_1\rightarrow 0}
\sinh{\pi (\l_2-\e_1)}\sum_{l_1=0}^{\infty} (-1)^{l_1}e^{-\d_1 l_1}
{Q(il_1+\e_1,\l_2)\over \l_2-il_1-\e_1}
\label{exn23}
\ee
Since $Q(il_1+\e_1,\l_2)$ is a polynomial on $l_1$ also again let us pick out 
some monomial from it
$$
l_1^{a'}Q'(\l_2)
$$
with an integer $a'\ge 0$ and a polynomial $Q'(\l_2)$.
Then the corresponding contribution into the expression (\ref{exn23}) for $J_2$
looks as follows
\be
\int_{C_{-1/2}}{d\l_2\over 2\pi i\,\sinh^2{\pi\l_2}} Q'(\l_2)
\lim_{\d_1\rightarrow 0^+} {d^{(2)}(\e_1)}_{\e_1\rightarrow 0}
\sinh{\pi (\l_2-\e_1)}\sum_{l_1=0}^{\infty} (-1)^{l_1}e^{-\d_1 l_1}
{l_1^{a'}\over \l_2-il_1-\e_1}
\label{cont1}
\ee
Since $\l_2=x_2-i/2$ with a real number $x_2\in (-\infty,\infty)$ 
the denominator
$$
{1\over \l_2-il_1-\e_1}\;=\;{i\over l_1+1/2+ix_2-i\e_1}
$$
and $Re(l_1+1/2+ix_2-i\e_1)\ge 1/2$ because $l_1\ge 0$. Therefore one can use an evident
integral representation
\be
{1\over\a}\;=\;\int_{0}^1{ds\over s}s^{\a}
\label{ints}
\ee
which is valid if $Re(\a)>0$.
Hence the sum in the expression (\ref{cont1}) is
\be
i\sum_{l_1=0}^{\infty}(-1)^{l_1}e^{-\d_1l_1}l_1^{a'}\int_0^1{ds\over s}
s^{l_1+1/2+ix_2-i\e_1}
\label{suml1}
\ee
Since for $\d_1>0$ the sum
$$
\sum_{l_1=0}^{N'}(-1)^{l_1}e^{-\d_1l_1}l_1^{a'}s^{l_1}\;=\;
(-{\partial\over\partial\d_1})^{a'}{1\,-\,(-e^{-\d_1}s)^{N'+1}\over 1\,+\,e^{-\d_1}s}
$$
where $N'$ is a positive integer
converges uniformly in $s\in [0,1]$ when $N'\rightarrow\infty$ then
one can interchange the sum over $l_1$ and the integration over $s$.
The result for (\ref{suml1}) is as follows
\be
i\int_0^1{ds\over s}s^{i \l_2 -i\e_1}
(-{\partial\over\partial\d_1})^{a'}{1\over 1\,+\,e^{-\d_1}s}\;=\;
i\int_0^1{ds\over s}s^{i \l_2 -i\e_1}
\sum_{k=1}^{a'+1}C(a',k){1\over (1\,+\,e^{-\d_1}s)^{k}}
\label{suml12}
\ee
where $C(a',k)$ are some rational coefficients.
So the contribution of $k$-th term into the expression (\ref{cont1}) is
\be
\int_{C_{-1/2}}{d\l_2\over 2\pi i\,\sinh^2{\pi\l_2}} Q'(\l_2)
\lim_{\d_1\rightarrow 0^+} {d^{(2)}(\e_1)}_{\e_1\rightarrow 0}
i\int_0^1{ds\over s}s^{i\l_2-i\e_1}C(a',k){1\over (1\,+\,e^{-\d_1}s)^{k}}
\label{cont2}
\ee
It is not very difficult to check that the function 
$$
Q'(\l_2)\sinh{\pi (\l_2-\e_1)}s^{i\l_2}
$$
satisfies the conditions (\ref{limR}) and (\ref{limN}) of the proposition 2 with $n=2$.
Moreover the integral
$$
\int_{C_{-1/2}}{d\l_2\over 2\pi i}
e^{i\d_2\l_2} {\sinh{\pi (\l_2-\e_1)}\over \sinh^2{\pi\l_2}} Q'(\l_2)\,s^{i\l_2}
$$
converges uniformly in $s\in [0,1]$ . Since it, actually, does not depend on $\d_1$ also
and $d^{(2)}(\e_1)$ given by (\ref{d2}) is the second derivative on $\e_1$
one can calculate the
integral over $\l_2$ first using the formulae (\ref{int1}-\ref{int2}) of the proposition 2
and then calculate the integral over $s$ and
take the limits on $\e_1$ and $\d_1$. 

Therefore we have shown that the formula (\ref{IS}) is correct for $n=2$.
Below we will use this formula for the concrete calculation of $P(2)$.

A generalization of our discussion to the arbitrary $n$ is straightforward and we will
not do it here.

The most efficient way of taking the integrals is as follows.
Fist we apply the formula (\ref{IS}) using either (\ref{int1}) or
(\ref{int2}) in such a way that the denominators like
$$
{1\over \l_2-\l_1} \quad\quad \mbox{becomes} \quad\quad {i\over l_1+l_2+i(\e_2-\e_1)}
$$
with $l_1\geq 0$ and $l_2\geq 1$.
In this case we shall not face a singularities like $1/(\e_2-\e_1)$
and the whole expression will be analytic on $\e_1,\ldots,\e_n$.
Hence, we can use the differential operator (\ref{Da}).
Then after using the formula like (\ref{ints}) one can get rid of all
such denominators and expand over $\e$-s until the order which still makes
a non-zero contribution after applying the differential operator (\ref{Da}).
It can be applied after
taking all summations in (\ref{F}).
As the last step one should take integrals on auxiliary variables like 
$s$ appeared above.

Below we shall also take all $\d_1,\ldots, \d_n$ to be zero at once implying
the limiting procedure $\lim_{\d_j\rightarrow 0^+}$ described above.

\vspace{0.3cm}

Now let us illustrate how the whole procedure 
works for a simple case $P(2)$
\be
P(2)\;=\;\pi^3\int_{C_{-1/2}}{d\l_1\over 2\pi i}
\int_{C_{-1/2}}{d\l_2\over 2\pi i}\>
{\sinh{\pi(\l_2-\l_1)}\over \sinh^2{\pi\l_1} \sinh^2{\pi\l_2}}\>
T_2(\l_1,\l_2)
\label{PP2}
\ee
In this case it is very simple to perform the first step, namely,
to get the "canonical" form (\ref{T_n^c}) described in the 
beginning of the Section because we do not need to reduce a
power of denominator in this case.
Indeed,
\be
T_2(\l_1,\l_2)\;=\;{(\l_1+i)\l_2\over\l_2-\l_1-i}\;=\;
\l_1+i\;+\;{(\l_1+i)^2\over\l_2-\l_1-i}\;\sim\;
\l_1\;-\;{\l_1^2\over\l_2-\l_1}
\label{T_2}
\ee
where we have used the property I and the formula
(\ref{l^m}) from the proposition 1 of the item IV.
Then using the formula (\ref{cor2}) of the corollary 2 for
$m=2,3$ one gets
$$
-\;{\l_1^2\over\l_2-\l_1}\;\sim\;-{i\over 3}\;
\biggl({3i^2\l_1+i^3\over\l_2-\l_1}\;+\;\l_2^2\;+\l_2(\l_1+i)\;+\;
(\l_1+i)^2\biggr)\;\sim\;
$$
$$
\sim\;{i\l_1\over\l_2-\l_1}\;
-\;{1\over 3}\,{1\over\l_2-\l_1}\;-\;{i\over 3}(i\l_1)\;\sim\;
{1\over 2}\,{1\over\l_2-\l_1}\;-\;{1\over 3}\,{1\over \l_2-\l_1}\;+\;
{1\over 3}
\l_1\;=\;{1\over 3}\l_1\;+\;{1\over 6}\,{1\over \l_2-\l_1}
$$
Substituting it to the formula (\ref{T_2}) we get
\be
T_2(\l_1,\l_2)\;\sim\;T_2^c(\l_1,\l_2)
\label{T_2^c}
\ee
where
\be
T_2^c(\l_1,\l_2)\;=\;{4\over 3}\,\l_1\;+\;{1\over 6}\,{1\over \l_2-\l_1}
\label{T_2^cdef}
\ee
and this is the "canonical" form for $T_2$ i.e. the polynomials
$P_0^{(2)}$ and $P_1^{(2)}$ from (\ref{T_n^c}-\ref{P_j}) are 
equal to $4/3\l_1$ and $1/6$ respectively.

Let us take the integral from the first term using the formula
(\ref{F}), our comments about the limiting procedure like in the
formulae (\ref{int1}-\ref{int2}) of the proposition 2
$$
J_0^{(2)}\;=\;\pi^3\int_{C_{-1/2}}{d\l_1\over 2\pi i}
\int_{C_{-1/2}}{d\l_2\over 2\pi i}\>
{\sinh{\pi(\l_2-\l_1)}\over \sinh^2{\pi\l_1} \sinh^2{\pi\l_2}}\>
{4\over 3}\,\l_1\;=\;D^{(2)}{4\over 3}\sum_{l_1=0}^{\infty}
(-1)^{l_1}(il_1+\e_1)\sum_{l_2=0}^{\infty}(-1)^{l_2}\;=\;
$$
\be
=\;D^{(2)}{4\over 3}(i\r(1)+\e_1\r(0))\r(0)\;=\;
D^{(2)}{4\over 3}(-{i\over 4}+{\e_1\over 2}){1\over 2}\;=\;
{({\partial\over\partial\e_1}-{\partial\over\partial\e_2})}_
{\e_1,\e_2\rightarrow 0}
{2\over 3}(-{i\over 4}+{\e_1\over 2})\;=\;{1\over 3}
\label{J02}
\ee
where we have used the formulae (\ref{ro5})
for $\r(b)$ given by (\ref{ro}) implying the limiting procedure
as it was explained above.

The second integral is treated as it was described in item VI
with the help of the integral representation (\ref{ints})
$$
J_1^{(2)}\;=\;\pi^3\int_{C_{-1/2}}{d\l_1\over 2\pi i}
\int_{C_{-1/2}}{d\l_2\over 2\pi i}\>
{\sinh{\pi(\l_2-\l_1)}\over \sinh^2{\pi\l_1} \sinh^2{\pi\l_2}}\>
{1\over 6}\;{1\over \l_2-\l_1}\,=\,
$$
$$
=\;D^{(2)}\>{(-1)\over 6}\sum_{l_1=1}^{\infty}\sum_{l_2=0}^{\infty}
\>{(-1)^{l_1+l_2}\over il_2+il_1+\e_2-\e_1}\;=\;
D^{(2)}\>{i\over 6}\sum_{l_1=1}^{\infty}\sum_{l_2=0}^{\infty}
\>{(-1)^{l_1+l_2}\over l_1+l_2+i(\e_1-\e_2)}\;=\;
$$
$$
=\;D^{(2)}\>{i\over 6}\int_0^1{ds\over s}
\sum_{l_1=1}^{\infty}\sum_{l_2=0}^{\infty}\>
(-1)^{l_1+l_2} s^{l_1+l_2+i(\e_1-\e_2)}\;=\;
D^{(2)}\>{(-i)\over 6}\int_0^1 ds\;
{s^{i(\e_1-\e_2)}\over (1+s)^2}\;=\;
$$
\be
=\;{({\partial\over\partial\e_1}-{\partial\over\partial\e_2})}_
{\e_1,\e_2\rightarrow 0}
{(-i)\over 6}\int_0^1 ds\;
{s^{i(\e_1-\e_2)}\over (1+s)^2}\;=\;
{1\over 3}\int_0^1 ds {\ln{s}\over (1+s)^2}\;=\;-{\ln{2}\over 3}
\label{J12}
\ee
Summing up two answers (\ref{J02}) and (\ref{J12}) we get the result
\be
P(2)\;=\;J_0^{(2)}\;+\;J_1^{(2)}\;=\;{1\over 3}\;-\;{\ln(2)\over 3}.
\label{P2res}
\ee
which coincides with the formula (\ref{P2}).

In the Appendices A and B we shall derive the formulae (\ref{P3}) and
(\ref{P4}) for $P(3)$ and $P(4)$ respectively.
In the end of this Section let us note that both results (\ref{P3})
and (\ref{P4}) are expressed in terms of the logarithmic function 
and the Riemann zeta function of odd arguments and do not depend 
on polylogarithms like, for example, $\mbox{Li}_4(1/2)$.
All coefficients before those functions in (\ref{P1}-\ref{P4}) 
are rational. Also they do not contain any powers of $\pi$ which 
could be considered as Riemann zeta functions of even arguments,
see the formula (\ref{zeta2n}) from the Introduction.

{\bf Our conjecture is that the final answer
for any $P(n)$ will also be expressed in terms of logarithm
$\ln{2}$ and Riemann zeta functions $\zeta(k)$ with odd integers $k$
and with rational coefficients.}

In fact, this conjecture is intimately connected with our hypothesis
from the beginning of this Section that the function
$T_n$ (\ref{T_n}) can be reduced
to the "canonical" form. Looking at the "canonical" form (\ref{T_n^c})
one can conclude that only Riemann zeta functions and their products
can enter into the final answer because all the denominators in
the r.h.s. of (\ref{T_n^c}) are  split out. It means that after
applying the formula (\ref{F}) the multiple summation can be performed
by pairs, say, $\sum_{l_{2k-1}}$ and $\sum_{l_{2k}}$. Each pair of these
summations results in some combination of zeta functions.

\section{Conclusion}

We want to emphasize an interesting connection between integrable and 
disordered models.
In order to describe correlations in integrable models one can use 
integrable integral 
operators \cite{iiks}. On the other hand   Tracy and Widom showed that 
in matrix models the distribution of eigenvalues and level spacing can 
be described
by the   integral operators, belonging to the same integrable class 
\cite{tw}.

Our current  work supports this link between integrable models and
chaotic models. Riemann zeta function  appears in the
description of both kind of models.

Let us repeat that the main result of this paper is the calculation
of $P(3)$ and $P(4)$ (\ref{P3}-\ref{P4})
by means of the multi-integral representation (\ref{intPn}).
The fact that only the logarithm $\ln{2}$ and Riemann zeta
function with odd arguments participate in the answers for
$P(1),\ldots, P(4)$ and with rational coefficients before these
functions
allows us to suppose that this is the general
property of $P(n)$. One could compare the calculation of $P(n)$ with
the many-loop calculation of the self-energy diagrams in the
renormalizable
quantum field theory which can also be expressed in terms of $\zeta$
functions
of odd arguments \cite{KR} .

Unfortunately, so far we have not got even a conjecture for $P(n)$ but
we believe that it is not an unsolvable problem. May be already
after calculation of $P(5)$ one could guess the right formula
for a generic case $P(n)$.
It would give an answer to the question discussed in the previous
section, namely, the question about the law of decay of $P(n)$
when $n$ tends to infinity.

Also it would be interesting to generalize above results to
the XXZ spin chain. Some interesting conjectures were recently
invented
by Razumov and Stroganov \cite{RS}
for the special case of the XXZ model
with $\Delta=-1/2$.
These conjectures would be supported if
it were possible to get $P(n)$ from the general integral
representation obtained by the RIMS group \cite{JMMN}.

\section{Acknowledgements}

The authors would like to thank  B. McCoy,  A.~Razumov,
M.~Shiroishi,  Yu.~Stroganov, M.~Takahashi, L.Takhtajan  for useful discussions.
This research  has been supported by the NSF grant PHY-9988566
and by INTAS Grant no. 01-561.

\section{Appendix A}

Here we discuss in detail the calculation of $P(3)$ performed 
by means of the general procedure described above.

As was pointed out in the beginning of Section 2 the first step 
should be a reduction of the function $T_3(\l_1,\l_2,\l_3)$ 
to the form (\ref{T_n^c}) which we have called the "canonical" form.
In comparison with the case $P(2)$ here we should reduce the 
power of the denominator in $T_3(\l_1,\l_2,\l_3)$. To do this
we will use the formula (\ref{re1}) from the item II. Namely,
\be
T_3(\l_1,\l_2,\l_3)\;=\;{(\l_1+i)^2(\l_2+i)\l_2\l_3^2\over
(\l_2-\l_1-i)(\l_3-\l_1-i)(\l_3-\l_2-i)}\;=\;
I_1^{(3)}\;+\;I_2^{(3)}\;+\;I_3^{(3)}
\label{T3}
\ee
where
\be
I_1^{(3)}\;=\;i{(\l_1+i)^2(\l_2+i)\l_2\l_3^2\over
(\l_3-\l_1-i)(\l_3-\l_2-i)},
\label{I31}
\ee
\be
I_2^{(3)}\;=\;i{(\l_1+i)^2(\l_2+i)\l_2\l_3^2\over
(\l_2-\l_1-i)(\l_3-\l_1-i)},
\label{I32}
\ee
\be
I_3^{(3)}\;=\;-i{(\l_1+i)^2(\l_2+i)\l_2\l_3^2\over
(\l_2-\l_1-i)(\l_3-\l_2-i)},
\label{I33}
\ee
Due to the $1\leftrightarrow 2$ symmetry 
of the denominator the first term $I_1^{(3)}$ can be simplified 
as follows
$$
I_1^{(3)}\;\sim\;-{(\l_1+i)(\l_2+i)\l_2\l_3^2\over
(\l_3-\l_1-i)(\l_3-\l_2-i)}\;\sim\;
{(\l_1+i)(\l_2+i)\l_3^2\over
\l_3-\l_1-i}\;=\;
$$
\be
=\;(\l_1+i)(\l_2+i)(\l_1+\l_3+i)\;+\;
{(\l_1+i)^3(\l_2+i)\over\l_3-\l_1-i}\;\sim\;
\l_1^2\l_2\;-\;{(\l_1+i)^3(\l_3+i)\over\l_2-\l_1-i}
\label{II31}
\ee
The denominator of the second term (\ref{I32}) has the symmetry
under the transposition $2\leftrightarrow 3$. Therefore it can also
be simplified
$$
I_2^{(3)}\;\sim\;-{(\l_1+i)^2\l_2\l_3^2\over
(\l_2-\l_1-i)(\l_3-\l_1-i)}\;\sim\;
{(\l_1+i)^2\l_2\l_3\over
\l_2-\l_1-i}\;=\;
$$
\be
=\;-(\l_1+i)^2\l_3\;-\;
{(\l_1+i)^3\l_3\over\l_2-\l_1-i}\;\sim\;
\l_1^2\l_2\;-\;{(\l_1+i)^3\l_3\over\l_2-\l_1-i}
\label{II32}
\ee
The third term (\ref{I33}) is treated as follows
$$
I_3^{(3)}\;=\;
-i(\l_1+i)^2(\l_2+i)(\l_3+\l_2+i)\;-
$$
$$
-\;i{(\l_1+i)^2(\l_2+i)^3\over\l_3-\l_2-i}\;+\;
i{(\l_1+i)^3\l_3^2\over\l_2-\l_1-i}\;-\;
i{(\l_1+i)^3\l_3^3\over(\l_2-\l_1-i)(\l_3-\l_2-i)}\;\sim
$$
\be
\sim\;-\l_1^2\l_2\;-\;{(\l_1+i)^3(\l_3+i)^2\over\l_2-\l_1-i}\;+\;
i{(\l_1+i)^3\l_3^2\over\l_2-\l_1-i}\;-\;
i{(\l_1+i)^3\l_3^3\over(\l_2-\l_1-i)(\l_3-\l_2-i)}
\label{II33}
\ee
Now adding up all the three terms together we get
\be
T_3(\l_1,\l_2,\l_3)\;\sim\;\l_1^2\l_2\;-\;
i{(\l_1+i)^3\l_3^3\over(\l_2-\l_1-i)(\l_3-\l_2-i)}\;\sim\;
-\l_2\l_3^2\;-\;
i{(\l_1+i)^3\l_3^3\over(\l_2-\l_1-i)(\l_3-\l_2-i)}.
\label{T3a}
\ee
Let us note that up to this moment we have used only the symmetry
property (\ref{transp}) from the item I, the formula (\ref{re1})
from the item II and a simple algebra.

Now we would like to use the formula (\ref{l^m}) of the 
proposition 1 for $m=3$ and again apply the transposition formula
(\ref{transp})
$$
T_3(\l_1,\l_2,\l_3)\;\sim\;
-\l_2\l_3^2\;-\;
i{\l_1^3(\l_3+i)^3\over(\l_2-\l_1)(\l_3-\l_2)}\;\sim\;
$$
$$
\sim\;-\l_2\l_3^2\;+\;
i{(\l_1+i)^3\l_3^3\over(\l_2-\l_1)(\l_3-\l_2)}\;=\;
-\l_2\l_3^2\;+\;
i{\l_1^3\l_3^3+3i\l_1^2\l_3^3-3\l_1\l_3^3-i\l_3^3\over
(\l_2-\l_1)(\l_3-\l_2)}\;\sim\;
$$
$$
\sim\;-\l_2\l_3^2\;-\;
{3\l_1^2\l_3^2+3i\l_1\l_3(\l_3+\l_2)-\l_3^2-\l_3\l_2-\l_2^2\over
\l_2-\l_1}\;\sim\;
$$
\be
\sim\;-\l_2\l_3^2\;-\;
{3\l_1^2\l_3^2+3i\l_1\l_3^2+3i\l_1^2\l_3-\l_3^2-\l_3\l_1-\l_1^2\over
\l_2-\l_1}
\label{T3b}
\ee
Now we can reduce the power of $\l_1$ in the numerator of the second
term (\ref{T3b}) by applying the formula (\ref{cor2}) of the
corollary 2 from the item IV. Doing this we finally get the 
"canonical form" (\ref{T_n^c}-\ref{P_j}) of $T_3$
\be
T_3(\l_1,\l_2,\l_3)\;\sim\;T_3^c(\l_1,\l_2,\l_3)\;=\;
P_0^{(3)}\;+\;{P_1^{(3)}\over\l_2-\l_1}
\label{T_3^c}
\ee
where the polynomials $P_0^{(3)}$ and $P_1^{(3)}$ are as follows
\be
P_0^{(3)}\;=\;-2\l_2\l_3^2,\quad
P_1^{(3)}\;=\;{1\over 3}\;-\;i\l_1\;-\;i\l_3\;-\;2\l_1\l_3
\label{P013}
\ee
Let us note that if we express variables $\l_j$ through the real
variables $x_j$ via $\l_j=x_j-i/2$ in order to get
the polynomials $\tilde P_j^{(3)}$ ( see the formula
(\ref{P_ja})) we get especially simple formulae, namely,
\be
\tilde P_0^{(3)}\;=\;-2x_2x_3^2,\quad
\tilde P_1^{(3)}\;=\;-{1\over 6}\;-\;2x_1x_3.
\label{P013a}
\ee
So, the function 
\be
\tilde T_3^c(x_1,x_2,x_3)\;=\;
\tilde P_0^{(3)}\;+\;{\tilde P_1^{(3)}\over x_2-x_1}
\label{tildeT_3^c}
\ee
is odd i.e. 
$$
\tilde T_3^c(-x_1,-x_2,-x_3)\;=\;-\tilde T_3^c(x_1,x_2,x_3)
$$
as it should be according to the formula (\ref{tildeT_n^ca}) 
from the beginning of Section 2.

Now we are ready to calculate the integral in order to get
the result for $P(3)$
\be
P(3)\;=\;\prod_{j=1}^3\int_{C_{-1/2}}{d\l_j\over 2\pi i}\>
U_3(\l_1,\l_2,\l_3)\>T_3(\l_1,\l_2,\l_3)\;=\;
J_0^{(3)}\;+\;J_1^{(3)}
\label{P3a}
\ee
where
\be
J_0^{(3)}=\prod_{j=1}^3\int_{C_{-1/2}}{d\l_j\over 2\pi i}\>
U_3(\l_1,\l_2,\l_3)\>P_0^{(3)},\quad
J_1^{(3)}=\prod_{j=1}^3\int_{C_{-1/2}}{d\l_j\over 2\pi i}\>
U_3(\l_1,\l_2,\l_3)\>
{P_1^{(3)}\over\l_2-\l_1}
\label{J301}
\ee
Using the formulae (\ref{P013}),(\ref{F}),(\ref{Da}),
(\ref{ro5}) we get
$$
J_0^{(3)}\;=\;\prod_{j=1}^3\int_{C_{-1/2}}{d\l_j\over 2\pi i}\>
U_3(\l_1,\l_2,\l_3)\>(-2)\l_2\l_3^2\;=\;
$$
$$
=\;D^{(3)}\sum_{l_1=0}^{\infty}(-1)^{l_1}
\sum_{l_2=0}^{\infty}(-1)^{l_2}\sum_{l_3=0}^{\infty}(-1)^{l_3}
(-2)(i l_2+\e_2)(i l_3+\e_3)^2\;=
$$
\be
=\;D^{(3)}\sum_{l_1=0}^{\infty}(-1)^{l_1}
\sum_{l_2=0}^{\infty}(-1)^{l_2}\sum_{l_3=0}^{\infty}(-1)^{l_3}
(-2)\e_2\e_3^2\;=\;
{1\over 8}\prod_{1\le k<j\le 3}
{\Bigl({\partial\over\partial\e_k}-
{\partial\over\partial\e_j}\Bigr)}_
{\vec\e\rightarrow 0}
(-2)\e_2\e_3^2\;=\;{1\over 4}
\label{J30}
\ee

To calculate the second term $J_1^{(3)}$ we should also use
the integral representation (\ref{ints})
$$
J_1^{(3)}\;=\;\prod_{j=1}^3\int_{C_{-1/2}}{d\l_j\over 2\pi i}\>
U_3(\l_1,\l_2,\l_3)\>{{1\over 3}\;-\;i\l_1\;-\;i\l_3\;-\;2\l_1\l_3
\over \l_2-\l_1}\;=\;
$$
$$
=-D^{(3)}\sum_{l_1=1}^{\infty}(-1)^{l_1}
\sum_{l_2=0}^{\infty}(-1)^{l_2}\sum_{l_3=0}^{\infty}(-1)^{l_3}
{{1\over 3}\;-\;i(-il_1+\e_1)\;-\;i(il_3+\e_3)\;-\;
2(-il_1+\e_1)(il_3+\e_3)
\over il_2+il_1+\e_2-\e_1}=
$$
$$
=\;i\;D^{(3)}\sum_{l_1=1}^{\infty}(-1)^{l_1}
\sum_{l_2=0}^{\infty}(-1)^{l_2}\sum_{l_3=0}^{\infty}(-1)^{l_3}
{{1\over 3}\;-\;l_1\;+\;l_3\;-i\e_1\;-\;i\e_3\;-\;
2(l_1+i\e_1)(l_3-i\e_3)
\over l_1+l_2+i(\e_1-\e_2)}\;=\;
$$
$$
=\;i\;D^{(3)}\sum_{l_1=1}^{\infty}(-1)^{l_1}
\sum_{l_2=0}^{\infty}(-1)^{l_2}
{-{1\over 12}\;-\;i\e_3/2\;+l_1i\e_3\;-\;\e_1\e_3
\over l_1+l_2+i(\e_1-\e_2)}\;=\;
$$
$$
=\;i\;D^{(3)}\int_0^1{ds\over s}s^{i(\e_1-\e_2)}
\sum_{l_1=1}^{\infty}\sum_{l_2=0}^{\infty}
(-{1\over 12}\;-\;i\e_3/2\;+l_1i\e_3\;-\;\e_1\e_3)(-s)^{l_1+l_2}
\;=\;
$$
$$
=\;i\;D^{(3)}\int_0^1{ds\over s}s^{i(\e_1-\e_2)}
\biggl(
{(-s)\over(1+s)^2}(-{1\over 12}\;-\;i\e_3/2\;-\;\e_1\e_3)\;+\;
{(-s)\over(1+s)^3}i\e_3
\biggr)\;=\;
$$
$$
=\;{i\over 2}\prod_{1\le k<j\le 3}
{\Bigl({\partial\over\partial\e_k}-{\partial\over\partial\e_j}\Bigr)}_
{\vec\e\rightarrow 0}
\int_0^1{ds\over (1+s)^2}
\biggl(
{1\over 12}\;(-3\e_1^2\e_2\;+\;3\e_1\e_2^2)\;(-i)\;{\ln^3{s}\over 6}\;+
\;i\ln{s}\;\e_1^2\e_3\biggr)\;=\;
$$
\be
=\;\int_0^1ds{\ln{s}\over (1+s)^2}\;-\;
{1\over 12}\;\int_0^1ds{\ln^3{s}\over (1+s)^2}\;=\;
-\ln{2}\;+\;{3\over 8}\zeta(3)
\label{J31}
\ee
Summing up $J_0^{(3)}$ and $J_1^{(3)}$ we get the final answer (\ref{P3})
\be
P(3)\;=\;J_0^{(3)}\;+\;J_1^{(3)}\;=\;
{1\over 4}\;-\;\ln{2}\;+\;{3\over 8}\zeta(3).
\label{P3b}
\ee

\section{Appendix B}

Here we will calculate $P(4)$. This case is much more complicated
than the previous ones. But using our general technique we will
try to simplify our discussion as much as possible.

As above the first step of our scheme is to get the "canonical"
form for the function\\
$T_4(\l_1,\l_2,\l_3,\l_4)$.

But before start doing this let us list some useful formulae
which can be derived from the proposition 1 and corollary 2
in the case $n=4$
\be
{\l_1^3\over\l_2-\l_1}\;\sim\;{-{3\over 2}i\l_1^2\;+\;\l_1\;+\;{i\over 4}
\over\l_2-\l_1}\;+\;{1\over 2}\l_2^2\;+\;{i\over 2}\l_2
\label{l^3}
\ee

\be
{\l_1^4\over\l_2-\l_1}\;\sim\;{-\l_1^2\;-\;i\l_1\;+\;{3\over 10}
\over\l_2-\l_1}\;+\;{3\over 5}\l_2^3\;+\;{1\over 5}\l_1\l_2^2\;+\;
{2\over 5}\l_2
\label{l^4}
\ee

\be
{\l_1^5\over\l_2-\l_1}\;\sim\;{-{\l_1\over 6}\;-\;{i\over 12}
\over\l_2-\l_1}\;+\;{1\over 3}\l_1\l_2^3\;-\;{4\over 3}i\l_2^3\;+\;
{13\over 6}\l_2^2\;+\;{4\over 3}i\l_2\;-\;{1\over 3}
\label{l^5}
\ee

$$
{\l_1^6\over\l_2-\l_1}\;\sim\;{-{1\over 2}\l_1^2\;
-\;{1\over 2}i\l_1\;-\;{1\over 7}\over\l_2-\l_1}\;+\;
$$
\be
+\;{1\over 7}\l_1^2\l_2^3\;-\;{6\over 7}i\l_1\l_2^3\;+
\;{2\over 7}\l_2^3\;+\;{10\over 7}\l_1\l_2^2\;-\;{25\over 14}i\l_2^2+
{31\over 14}\l_2
\label{l^6}
\ee

\be
{(\l_1+i)^3\over\l_2-\l_1}\;\sim\;
{{3\over 2}i\l_1^2\;-\;2\l_1\;-\;{3\over 4}i
\over\l_2-\l_1}\;+\;{1\over 2}\l_2^2\;+\;{i\over 2}\l_2
\label{lpi^3}
\ee

\be
{(\l_1+i)^4\over\l_2-\l_1}\;\sim\;{-\l_1^2\;-\;i\l_1\;+{3\over 10}
\over\l_2-\l_1}\;+\;{3\over 5}\l_2^3\;+\;{1\over 5}\l_1\l_2^2\;+\;
2i\l_2^2\;-\;{8\over 5}\l_2
\label{lpi^4}
\ee

\be
{(\l_1+i)^5\over\l_2-\l_1}\;\sim\;{-{\l_1\over 6}\;-\;{i\over 12}
\over\l_2-\l_1}\;+\;{1\over 3}\l_1\l_2^3\;+\;{5\over 3}i\l_2^3\;+\;
i\l_1\l_2^2\;-\;
{17\over 6}\l_2^2\;-\;{5\over 3}i\l_2\;-\;{1\over 3}
\label{lpi^5}
\ee
Here we have omitted an arbitrary function $g(\l_3,\l_4)$ 
that can be multiplied on the r.h.s. and l.h.s. of 
(\ref{l^3}-\ref{lpi^5}) as in the formula (\ref{cor2}).

Now we can start our derivation.
Fortunately, due to the formula (\ref{ratio}) of the observation 
III which in this case looks as follows
\be
{T_4(\l_1,\l_2,\l_3,\l_4)\over T_3(\l_1,\l_2,\l_3)}
\;=\;{(\l_1+i)(\l_2+i)(\l_3+i)\l_4^3\over
(\l_4-\l_1-i)(\l_4-\l_2-i)(\l_4-\l_3-i)}
\label{ratio43}
\ee
and has a symmetry under any permutation of the variables
$\l_1,\l_2,\l_3$ we can use the result (\ref{T3a}) that was obtained 
by means of the symmetry and simple algebra.
So, after the application of the formulae (\ref{T3a}) 
and (\ref{ratio43}) we can start with the following expression
$$
T_4(\l_1,\l_2,\l_3,\l_4)\;\sim\;
-{(\l_1+i)(\l_2+i)\l_2(\l_3+i)\l_3^2\l_4^3\over
(\l_4-\l_1-i)(\l_4-\l_2-i)(\l_4-\l_3-i)}\;-\;
$$
\be
-\;i{(\l_1+i)^4(\l_2+i)(\l_3+i)\l_3^3\l_4^3\over
(\l_2-\l_1-i)(\l_3-\l_2-i)(\l_4-\l_1-i)(\l_4-\l_2-i)(\l_4-\l_3-i)}
\label{T4a}
\ee
The denominator of the first term in the r.h.s. of (\ref{T4a}) 
is symmetric under permutation of $\l_1,\l_2,\l_3$. Therefore
it can be simplified as follows
$$
-{(\l_1+i)(\l_2+i)\l_2(\l_3+i)\l_3^2\l_4^3\over
(\l_4-\l_1-i)(\l_4-\l_2-i)(\l_4-\l_3-i)}\;\sim\;
{(\l_1+i)(\l_2+i)\l_2(\l_3+i)(\l_3+\l_4+i)\l_4^3\over
(\l_4-\l_1-i)(\l_4-\l_2-i)}\;\sim\;
$$
$$
\sim\;-{(\l_1+i)(\l_2+i)(\l_3+i)(\l_3+\l_4+i)\l_4^3\over
\l_4-\l_1-i}\;\sim\;
-{(\l_1+i)(\l_2+i)(\l_3+i)\l_3\l_4^3\over
\l_4-\l_1-i}\;=\;
$$
$$
=\;(\l_2+i)(\l_3+i)\l_3\l_4^3\;-\;
{(\l_2+i)(\l_3+i)\l_3\l_4^4\over\l_4-\l_1-i}\;\sim\;
\l_2\l_3^2\l_4^3\;-\;{\l_1^4(\l_3+i)(\l_4+i)\l_4\over
\l_2-\l_1+i}\;\sim
$$
$$
\sim\;\l_2\l_3^2\l_4^3\;-\;{\l_1^4(\l_3\l_4^2+i\l_4^2-\l_4)\over
\l_2-\l_1+i}\;\sim\;
\l_2\l_3^2\l_4^3\;+\;{(\l_1+i)^4(\l_3\l_4^2+i\l_4^2-\l_4)\over
\l_2-\l_1}\;\sim\;
$$
\be
\sim\;{8\over 5}\l_2\l_3^2\l_4^3\;-\;{(\l_1^2\;+\;i\l_1\;-\;{3\over 10})\>
(\l_3\l_4^2+i\l_4^2-\l_4)\over\l_2-\l_1}
\label{T4b}
\ee
The latter formula was obtained with the help of the formula
(\ref{lpi^4}). In fact, it is nothing but the "canonical" form
(\ref{T_n^c}) for the first term in (\ref{T4a}).

Now let us treat the second term. Using the formulae (\ref{re1})
and (\ref{re2}) we can write it down as follows
$$
-i{(\l_1+i)^4(\l_2+i)(\l_3+i)\l_3^3\l_4^3\over
(\l_2-\l_1-i)(\l_3-\l_2-i)(\l_4-\l_1-i)(\l_4-\l_2-i)(\l_4-\l_3-i)}\;=\;
$$
$$
\phantom{a}=\;-i(\l_1+i)^4(\l_2+i)(\l_3+i)\l_3^3\l_4^3\>\biggl\{
\phantom{a}\qquad\qquad\qquad\qquad\qquad\qquad\qquad\qquad
$$
$$
{1\over 2}{1\over(\l_3-\l_2-i)(\l_4-\l_1-i)(\l_4-\l_3-i)}\;-\;
{1\over 2}{1\over(\l_2-\l_1-i)(\l_4-\l_1-i)(\l_4-\l_3-i)}\;-
$$
$$
-\;{1\over 2}{1\over(\l_2-\l_1-i)(\l_3-\l_2-i)(\l_4-\l_1-i)}\;-\;
{1\over 2}{1\over(\l_2-\l_1-i)(\l_3-\l_2-i)(\l_4-\l_3-i)}\;-
$$
$$
-\;{1\over(\l_2-\l_1)(\l_3-\l_1)(\l_4-\l_1-i)}\;-\;
{1\over(\l_3-\l_2-i)(\l_4-\l_1-i)(\l_4-\l_2-i)}\;-
$$
$$
-\;{1\over(\l_3-\l_1)(\l_3-\l_2)(\l_4-\l_3-i)}\;+\;
{1\over(\l_2-\l_1-i)(\l_4-\l_2-i)(\l_4-\l_3-i)}\;+
$$
$$
+\;{1\over(\l_2-\l_1)(\l_3-\l_2)(\l_4-\l_2-i)}\;+\;
{1\over(\l_2-\l_1-i)(\l_3-\l_2-i)(\l_4-\l_2-i)}	\>\biggr\}
$$
Let us enumerate all ten terms as they enter here and make some
appropriate transformations for each of them
\be
I_1^{(4)}=
-{i (\l_1+i)^4(\l_2+i)(\l_3+i)\l_3^3\l_4^3
\over 2 (\l_3-\l_2-i)(\l_4-\l_1-i)(\l_4-\l_3-i)}
\begin{array}{c}
\quad\\
\quad\\
\sim\\
{\tiny
\begin{array}{c}
1\rightarrow 4\\
2\rightarrow 2\\
3\rightarrow 1\\
4\rightarrow 3
\end{array}}
\end{array}
-{i (\l_1+i)\l_1^3(\l_2+i)\l_3^3(\l_4+i)^4
\over 2 (\l_2-\l_1+i)(\l_3-\l_1-i)(\l_4-\l_3+i)}
\label{I41}
\ee
\be
I_2^{(4)}\;=\;
{i\over 2}{(\l_1+i)^4(\l_2+i)(\l_3+i)\l_3^3\l_4^3
\over(\l_2-\l_1-i)(\l_4-\l_1-i)(\l_4-\l_3-i)}\;
\begin{array}{c}
\quad\\
\sim\\
{\tiny
\begin{array}{c}
3\leftrightarrow 4
\end{array}}
\end{array}
{i\over 2}{(\l_1+i)^4(\l_2+i)\l_3^3(\l_4+i)\l_4^3
\over(\l_2-\l_1-i)(\l_3-\l_1-i)(\l_4-\l_3+i)}
\label{I42}
\ee
\be
I_3^{(4)}=
{i\over 2}{(\l_1+i)^4(\l_2+i)(\l_3+i)\l_3^3\l_4^3
\over(\l_2-\l_1-i)(\l_3-\l_2-i)(\l_4-\l_1-i)}
\begin{array}{c}
\quad\\
\quad\\
\sim\\
{\tiny
\begin{array}{c}
1\rightarrow 1\\
2\rightarrow 3\\
3\rightarrow 4\\
4\rightarrow 2
\end{array}}
\end{array}
{i\over 2}{(\l_1+i)^4\l_2^3(\l_3+i)(\l_4+i)\l_4^3
\over(\l_2-\l_1-i)(\l_3-\l_1-i)(\l_4-\l_3-i)}
\label{I43}
\ee
\be
I_4^{(4)}\;=\;
{i\over 2}{(\l_1+i)^4(\l_2+i)(\l_3+i)\l_3^3\l_4^3
\over(\l_2-\l_1-i)(\l_3-\l_2-i)(\l_4-\l_3-i)}\;
\begin{array}{c}
\quad\\
\sim\\
{\tiny
\begin{array}{c}
1\leftrightarrow 2
\end{array}}
\end{array}
{i\over 2}{(\l_1+i)(\l_2+i)^4(\l_3+i)\l_3^3\l_4^3
\over(\l_2-\l_1+i)(\l_3-\l_1-i)(\l_4-\l_3-i)}
\label{I44}
\ee
\be
I_5^{(4)}\;=\;
i{(\l_1+i)^4(\l_2+i)(\l_3+i)\l_3^3\l_4^3
\over(\l_2-\l_1)(\l_3-\l_1)(\l_4-\l_1-i)}
\label{I45}
\ee
\be
I_6^{(4)}=
i{(\l_1+i)^4(\l_2+i)(\l_3+i)\l_3^3\l_4^3
\over(\l_3-\l_2-i)(\l_4-\l_1-i)(\l_4-\l_2-i)}
\begin{array}{c}
\quad\\
\quad\\
\sim\\
{\tiny
\begin{array}{c}
1\rightarrow 4\\
2\rightarrow 1\\
3\rightarrow 2\\
4\rightarrow 3
\end{array}}
\end{array}
i{(\l_1+i)(\l_2+i)\l_2^3\l_3^3(\l_4+i)^4
\over(\l_2-\l_1-i)(\l_3-\l_1-i)(\l_4-\l_3+i)}
\label{I46}
\ee
\be
I_7^{(4)}\;=\;
i{(\l_1+i)^4(\l_2+i)(\l_3+i)\l_3^3\l_4^3
\over(\l_3-\l_1)(\l_3-\l_2)(\l_4-\l_3-i)}\;
\begin{array}{c}
\quad\\
\sim\\
{\tiny
\begin{array}{c}
1\leftrightarrow 3
\end{array}}
\end{array}
-i{(\l_1+i)\l_1^3(\l_2+i)(\l_3+i)^4\l_4^3
\over(\l_2-\l_1)(\l_3-\l_1)(\l_4-\l_1-i)}
\label{I47}
\ee
\be
I_8^{(4)}=
-{i(\l_1+i)^4(\l_2+i)(\l_3+i)\l_3^3\l_4^3
\over(\l_2-\l_1-i)(\l_4-\l_2-i)(\l_4-\l_3-i)}
\begin{array}{c}
\quad\\
\quad\\
\sim\\
{\tiny
\begin{array}{c}
1\leftrightarrow 2\\
3\leftrightarrow 4
\end{array}}
\end{array}
-{i(\l_1+i)(\l_2+i)^4\l_3^3(\l_4+i)\l_4^3
\over(\l_2-\l_1+i)(\l_3-\l_1-i)(\l_4-\l_3+i)}
\label{I48}
\ee
\be
I_9^{(4)}\;=\;
-i{(\l_1+i)^4(\l_2+i)(\l_3+i)\l_3^3\l_4^3
\over(\l_2-\l_1)(\l_3-\l_2)(\l_4-\l_2-i)}\;
\begin{array}{c}
\quad\\
\quad\\
\sim\\
{\tiny
\begin{array}{c}
1\rightarrow 3\\
2\rightarrow 1\\
3\rightarrow 2\\
4\rightarrow 4\\
\end{array}}
\end{array}
i{(\l_1+i)(\l_2+i)\l_2^3(\l_3+i)^4\l_4^3
\over(\l_2-\l_1)(\l_3-\l_1)(\l_4-\l_1-i)}
\label{I49}
\ee
\be
I_{10}^{(4)}=
-{i(\l_1+i)^4(\l_2+i)(\l_3+i)\l_3^3\l_4^3
\over(\l_2-\l_1-i)(\l_3-\l_2-i)(\l_4-\l_2-i)}
\begin{array}{c}
\quad\\
\sim\\
{\tiny
\begin{array}{c}
1\leftrightarrow 2
\end{array}}
\end{array}
-{i(\l_1+i)(\l_2+i)^4(\l_3+i)\l_3^3\l_4^3
\over(\l_2-\l_1+i)(\l_3-\l_1-i)(\l_4-\l_1-i)}
\label{I410}
\ee
Let us note that after these transformations $I_j^{(4)}$ have two kinds
of the denominators, namely, $I_5^{(4)},I_7^{(4)},I_9^{(4)},
I_{10}^{(4)}$ have denominators of the form 
$$
{1\over(\l_2-\l_1-ia)(\l_3-\l_1-ib)(\l_4-\l_1-ic)}
$$ 
with a set of integers
$a,b,c$, while the denominators of the rest of them are of the form
$$
{1\over(\l_2-\l_1-ia')(\l_3-\l_1-ib')(\l_4-\l_3-ic')}
$$ 
with some other set of integers $a',b',c'$.
Moreover, some
of the terms $I_j^{(4)}$ have just coinciding denominators like, for
example, $I_2^{(4)}$ and $I_6^{(4)}$. Nevertheless sometimes it will be 
more convenient to treat them separately.

Let us start with the first group which is easier to treat.
Since, the denominator of $I_5^{(4)}$ has the $2\leftrightarrow 3$ 
symmetry we can simplify it as follows
\be
I_5^{(4)}\;\sim\;i{(\l_1+i)^4(\l_2+i)(\l_3+i)(\l_3^2\,+\,\l_3\l_1
\,+\,\l_1^2)\l_4^3
\over(\l_2-\l_1)(\l_4-\l_1-i)}
\label{I45a}
\ee
where we have used the trivial identity
\be
\l_3^3\;=\;(\l_3-\l_1)(\l_3^2\,+\,\l_3\l_1\,+\,\l_1^2)\;+\;\l_1^3
\label{l3^3}
\ee
and the fact that the second term in the r.h.s. of (\ref{l3^3}) does
not contribute into $I_5^{(4)}$.

Then summing up (\ref{I47}) and (\ref{I49}) we get
$$
I_7^{(4)}\;+\;I_9^{(4)}\;\sim\;
i{(\l_1+i)(\l_2+i)(\l_1^2\,+\,\l_1\l_2\,+\,\l_2^2)(\l_3+i)^4\l_4^3
\over(\l_3-\l_1)(\l_4-\l_1-i)}
\begin{array}{c}
\quad\\
\sim\\
{\tiny
\begin{array}{c}
1\leftrightarrow 2
\end{array}}
\end{array}
$$
\be
\sim\;-i{(\l_1+i)(\l_2+i)^4(\l_3+i)
(\l_1^2\,+\,\l_1\l_3\,+\,\l_3^2)\l_4^3
\over(\l_2-\l_1)(\l_4-\l_1-i)}
\label{I479}
\ee
Now adding the r.h.s. of (\ref{I45a}) to (\ref{I479}) we get
$$
I_5^{(4)}\;+\;I_7^{(4)}\;+\;I_9^{(4)}\;\sim\;
$$
$$
\sim\;-i{(\l_1+i)(\l_2+i)(\l_3+i)
((\l_1+i)^2\,+\,(\l_1+i)(\l_2+i)\,+\,(\l_2+i)^2)
(\l_3^2+\l_3\l_1+\l_1^2)\l_4^3
\over\l_4-\l_1-i}\;\sim\;
$$
$$
\sim\;3\;{(\l_1+i)(\l_2+i)(\l_3+i)
(\l_1\,+\,\l_2\,+\,i)(\l_3^2+\l_3\l_1+\l_1^2)\l_4^3
\over\l_4-\l_1-i}\;=\;
$$
$$
=\;-3\;(\l_2+i)(\l_3+i)(\l_1+\l_2+i)(\l_3^2+\l_3\l_1+\l_1^2)\l_4^3\;+\;
$$
$$
\;+\;3\;{(\l_2+i)(\l_3+i)
(\l_1\,+\,\l_2\,+\,i)(\l_3^2+\l_3\l_1+\l_1^2)\l_4^4
\over\l_4-\l_1-i}\;\sim\;
$$
$$
\sim\;3\;{(\l_2+i)(\l_3+i)
(\l_1\,+\,\l_2\,+\,i)(\l_3^2+\l_3\l_1+\l_1^2)\l_4^4
\over\l_4-\l_1-i}\;\sim\;
$$
$$
\sim\;3\;{(\l_1+i)(\l_2+i)(\l_3+i)^2\l_3\l_4^4
\over\l_4-\l_1-i}\;+\;
3\;{(\l_2+i)(\l_3+i)\l_3(\l_2\l_3+1)\l_4^4
\over\l_4-\l_1-i}\;\sim\;
$$
$$
\sim\;-3\l_2\l_3^2\l_4^3\;+\;
3\;{(\l_2+i)(\l_3+i)^2\l_3\l_4^5
\over\l_4-\l_1-i}\;+\;
3\;{(\l_2+i)(\l_3+i)\l_3(\l_2\l_3+1)\l_4^4
\over\l_4-\l_1-i}\;\sim\;
$$
$$
\sim\;-3\l_2\l_3^2\l_4^3\;+\;
3\;{(\l_2+i)(\l_3+i)^2\l_3(\l_4+i)^5
\over\l_4-\l_1}\;+\;
3\;{(\l_2+i)(\l_3+i)\l_3(\l_2\l_3+1)(\l_4+i)^4
\over\l_4-\l_1}
\begin{array}{c}
\quad\\
\quad\\
\sim\\
{\tiny
\begin{array}{c}
1\rightarrow 2\\
2\rightarrow 3\\
3\rightarrow 4\\
4\rightarrow 1\\
\end{array}}
\end{array}
$$
$$
\sim\;-3\l_2\l_3^2\l_4^3\;-\;
3\;{(\l_1+i)^5(\l_3+i)(\l_4+i)^2\l_4
\over\l_2-\l_1}\;-\;
3\;{(\l_1+i)^4(\l_3+i)(\l_4+i)\l_4(\l_3\l_4+1)
\over\l_2-\l_1}\;\sim\;
$$
$$
\sim\;-{17\over 10}\l_2\l_3^2\l_4^3\;+\;
{({\l_1\over 2}\,+\,{i\over 4})(\l_3+i)(\l_4+i)^2\l_4
\over\l_2-\l_1}\;+\;
$$
$$
+\;{(3\l_1^2\,+\,3i\l_1\,-\,{9\over 10})(\l_3+i)(\l_4+i)\l_4(\l_3\l_4+1)
\over\l_2-\l_1}\;\sim\;
-{17\over 10}\l_2\l_3^2\l_4^3\;+\;
$$
\be
+\;{
({\l_1\over 2}\,+\,{i\over 4})
(\l_3\l_4^3+i\l_4^3+2i\l_3\l_4^2-2\l_4^2-i\l_4)\:+\;
(3\l_1^2\,+\,3i\l_1\,-\,{9\over 10})
(\l_3^2\l_4^3+i\l_3\l_4^3+i\l_4^2-\l_4)
\over\l_2-\l_1}
\label{I579}
\ee
In fact, it is the "canonical" form.
To get this expression we have used symmetries, the
formula (\ref{l^m}) for $m=4$ and $m=5$
of the proposition 1 and relations
(\ref{lpi^4}) and (\ref{lpi^5}).

Now let us treat $I_{10}^{(4)}$
$$
I_{10}^{(4)}\;=\;
-i{(\l_1+i)(\l_2+i)^4(\l_3+i)\l_3^3\l_4^3
\over(\l_2-\l_1+i)(\l_3-\l_1-i)(\l_4-\l_1-i)}\;\sim\;
-i{(\l_1+i)(\l_2+i)^4\l_3^3\l_4^3
\over(\l_2-\l_1+i)(\l_4-\l_1-i)}\;=\;
$$
$$
=\;i{(\l_2+i)^4\l_3^3\l_4^3
\over\l_2-\l_1+i}\;-\;
i{(\l_2+i)^4\l_3^3\l_4^4
\over(\l_2-\l_1+i)(\l_4-\l_1-i)}\;\sim\;
-i{(\l_2+i)^4\l_3^3\l_4^4
\over(\l_2-\l_1+i)(\l_4-\l_1-i)}\;\sim\;
$$
$$
\sim\;-i{\l_2^4\l_3^3(\l_4+i)^4
\over(\l_2-\l_1)(\l_4-\l_1)}\;=\;
i{\l_3^3(\l_2^4\l_4^4+4i\l_2^3\l_4^4-6\l_2^2\l_4^4-4i\l_2\l_4^4+\l_4^4)
\over(\l_2-\l_1)(\l_4-\l_1)}\;\sim\;
$$
$$
\sim\;i{\l_3^3(4i\l_2^3\l_4^3-6\l_2^2\l_4^2(\l_4+\l_1)-
4i\l_2\l_4(\l_4^2+\l_4\l_1+\l_1^2)+\l_4^3+\l_4^2\l_1+\l_4\l_1^2+\l_1^3)
\over\l_2-\l_1}\;\sim\;
$$
$$
\sim\;i{\l_3^3(-6\l_1\l_2^2\l_4^2-
4i\l_1\l_2\l_4^2-4i\l_1^2\l_2\l_4+\l_1\l_4^2+\l_1^2\l_4+\l_1^3)
\over\l_2-\l_1}\;=\;
$$
$$
=\;i\l_3^3(-6\l_1^2\l_4^2-4i\l_1\l_4^2-4i\l_1^2\l_4)\;+\;
i{\l_3^3(-6\l_1^3\l_4^2-
4i\l_1^2\l_4^2-4i\l_1^3\l_4+\l_1\l_4^2+\l_1^2\l_4+\l_1^3)
\over\l_2-\l_1}\;\sim\;
$$
$$
\sim\;i{\l_3^3(\l_4^2(5i\l_1^2-5\l_1-{3\over 2}i)+
\l_4(-5\l_1^2+4i\l_1+1)-{3\over 2}i\l_1^2+\l_1+{i\over 4})
\over\l_2-\l_1}\;+\;
$$
$$
+\;\l_3^3(-3)\l_4^2i\l_2\;+\;i\l_3^3(-4)i\l_4
{1\over 2}\l_2^2
\;\sim\;
$$
\be
\sim\;-\l_2\l_3^2\l_4^3\;+\;
{(5\l_1^2+5i\l_1-{3\over 2})\l_3^2\l_4^3+
(5i\l_1^2-4\l_1-i)\l_3\l_4^3+(-{3\over 2}\l_1^2-i\l_1+{1\over 4})\l_4^3
\over\l_2-\l_1}
\label{I410a}
\ee
Here the symmetry, the formula (\ref{l^m}) for
$m=4$ and the relation (\ref{l^3}) were used.

Now we shall treat the other six terms $I_1^{(4)}, I_2^{(4)}, 
I_3^{(4)}, I_4^{(4)}, I_6^{(4)}$ and $I_8^{(4)}$ given by the
expressions (\ref{I41}),(\ref{I42}),(\ref{I43}),(\ref{I44}),
(\ref{I46}) and (\ref{I48}) respectively. Here we shall proceed in
two steps. As a first step we will reduce all these six terms to
a following form
\be
I_j^{(4)}\;\sim\;{A_j\over (\l_2-\l_1)(\l_3-\l_1)(\l_4-\l_3)}\;+\;
{B_j\over (\l_2-\l_1)(\l_3-\l_1)}\;+\;
{C_j\over (\l_2-\l_1)(\l_4-\l_3)}\;+\;
{D_j\over \l_2-\l_1}
\label{I123468}
\ee
where $j=1,2,3,4,6,8$ and $A_j,B_j,C_j,D_j$ are some polynomials.
Then we shall sum up all the six results and get the "canonical"
form for the sum obtained.

Let us start with the expression (\ref{I41}) for the term $I_1^{(4)}$ 
$$
I_1^{(4)}=
-{i\over 2}{(\l_1+i)\l_1^3(\l_2+i)\l_3^3(\l_4+i)^4
\over(\l_2-\l_1+i)(\l_3-\l_1-i)(\l_4-\l_3+i)}\;=\;
$$
$$
=\;-{i\over 2}{(\l_1^2+\l_1(\l_2+i)+(\l_2+i)^2)(\l_2+i)\l_3^3(\l_4+i)^4
\over\l_4-\l_3+i}\;+\;
$$
$$
+\;{i\over 2}{(\l_1^2+\l_1(\l_2+i)+(\l_2+i)^2)(\l_2+i)\l_3^4(\l_4+i)^4
\over(\l_3-\l_1-i)(\l_4-\l_3+i)}\;+\;
{i\over 2}{(\l_2+i)^4\l_3^3(\l_4+i)^3\over\l_2-\l_1+i}\;-\;
$$
$$
-\;{i\over 2}{(\l_2+i)^4\l_3^4(\l_4+i)^3\over
(\l_2-\l_1-i)(\l_3-\l_1+i)}\;-\;
{i\over 2}{(\l_1+i)(\l_2+i)^4\l_3^4(\l_4+i)^3\over
(\l_2-\l_1+i)(\l_3-\l_1-i)(\l_4-\l_3+i)}\;\sim\;
$$
$$
\sim\;{i\over 2}{(\l_1^2+\l_1(\l_2+i)+(\l_2+i)^2)(\l_2+i)\l_3^3\l_4^4
\over\l_4-\l_3}\;-\;
$$
$$
-\;{i\over 2}
{(\l_1^2+\l_1(\l_2+i)+(\l_2+i)^2)(\l_2+i)(\l_3+i)^4(\l_4+i)^4
\over(\l_3-\l_1)(\l_4-\l_3)}\;-\;
{i\over 2}{\l_2^4\l_3^3(\l_4+i)^3\over\l_2-\l_1}\;-\;
$$
$$
-\;{i\over 2}{\l_2^4(\l_3+i)^4(\l_4+i)^3\over
(\l_2-\l_1)(\l_3-\l_1)}\;-\;
{i\over 2}{(\l_1+i)\l_2^4(\l_3+i)^4(\l_4+i)^3\over
(\l_2-\l_1)(\l_3-\l_1)(\l_4-\l_3)}\;\sim\;
$$
$$
\sim\;{i\over 2}{\l_1^3\l_2^4(\l_3^2+\l_3(\l_4+i)+(\l_4+i)^2)(\l_4+i)
\over\l_2-\l_1}\;+\;
$$
$$
+\;{i\over 2}
{(\l_1+i)^4(\l_2+i)^4(\l_3^2+\l_3(\l_4+i)+(\l_4+i)^2)(\l_4+i)
\over(\l_2-\l_1)(\l_3-\l_1)}\;-\;
{i\over 2}{\l_2^4\l_3^3(\l_4+i)^3\over\l_2-\l_1}\;-\;
$$
\be
-\;{i\over 2}{\l_2^4(\l_3+i)^4(\l_4+i)^3\over
(\l_2-\l_1)(\l_3-\l_1)}\;-\;
{i\over 2}{(\l_1+i)\l_2^4(\l_3+i)^4(\l_4+i)^3\over
(\l_2-\l_1)(\l_3-\l_1)(\l_4-\l_3)}
\label{I41a}
\ee
where we have used symmetries and the formula (\ref{l^m}) with $m=4$.

For the next term $I_2^{(4)}$ we start with the expression
(\ref{I42})
$$
I_2^{(4)}\;\sim\;
{i\over 2}{(\l_1+i)^4(\l_2+i)\l_3^3(\l_4+i)\l_4^3
\over(\l_2-\l_1-i)(\l_3-\l_1-i)(\l_4-\l_3+i)}\;=\;
$$
$$
=\;-{i\over 2}{(\l_1+i)^4(\l_2+i)
(\l_3^2+\l_3(\l_4+i)+(\l_4+i)^2)(\l_4+i)\l_4^3
\over(\l_2-\l_1-i)(\l_3-\l_1-i)}\;+\;
$$
$$
+\;{i\over 2}{(\l_1+i)^4(\l_2+i)(\l_4+i)^4\l_4^3
\over(\l_2-\l_1-i)(\l_3-\l_1-i)(\l_4-\l_3+i)}\;\sim\;
$$
\be
\sim\;{i\over 2}{\l_1^4(\l_2+i)
(\l_3^2+\l_3(\l_4+i)+(\l_4+i)^2)(\l_4+i)\l_4^3
\over(\l_2-\l_1)(\l_3-\l_1)}\;+\;
{i\over 2}{\l_1^4(\l_2+i)\l_4^4(\l_4-i)^3
\over(\l_2-\l_1)(\l_3-\l_1)(\l_4-\l_3)}
\label{I42a}
\ee

The term $I_3^{(4)}$ given by (\ref{I43}) is also simple to treat
$$
I_3^{(4)}=
{i\over 2}{(\l_1+i)^4\l_2^3(\l_3+i)(\l_4+i)\l_4^3
\over(\l_2-\l_1-i)(\l_3-\l_1-i)(\l_4-\l_3-i)}\;=\;
$$
$$
=\;-{i\over 2}{(\l_1+i)^4\l_2^3(\l_4+i)\l_4^3
\over(\l_2-\l_1-i)(\l_3-\l_1-i)}\;+\;
{i\over 2}{(\l_1+i)^4\l_2^3(\l_4+i)\l_4^4
\over(\l_2-\l_1-i)(\l_3-\l_1-i)(\l_4-\l_3-i)}\;\sim\;
$$
\be
\sim\;{i\over 2}{\l_1^4\l_2^3(\l_4+i)\l_4^3
\over(\l_2-\l_1)(\l_3-\l_1)}\;+\;
{i\over 2}{\l_1^4\l_2^3(\l_4+2i)(\l_4+i)^4
\over(\l_2-\l_1)(\l_3-\l_1)(\l_4-\l_3)}
\label{I43a}
\ee

The fourth term $I_4^{(4)}$ (\ref{I44}) 
demands more work in order to reduce
it to the form (\ref{I123468})
$$
I_4^{(4)}\;\sim\;
{i\over 2}{(\l_1+i)(\l_2+i)^4(\l_3+i)\l_3^3\l_4^3
\over(\l_2-\l_1+i)(\l_3-\l_1-i)(\l_4-\l_3-i)}\;=\;
$$
$$
=\;-{i\over 2}{(\l_1+i)(\l_2+i)^4\l_3^3\l_4^3
\over(\l_2-\l_1+i)(\l_3-\l_1-i)}\;+\;
{i\over 2}{(\l_1+i)(\l_2+i)^4\l_3^3\l_4^4
\over(\l_2-\l_1+i)(\l_3-\l_1-i)(\l_4-\l_3-i)}\;=\;
$$
$$
=\;{i\over 2}{(\l_2+i)^4\l_3^3\l_4^3
\over(\l_2-\l_1+i)}\;-\;
{i\over 2}{(\l_2+i)^4\l_3^4\l_4^3
\over(\l_2-\l_1+i)(\l_3-\l_1-i)}\;+\;
$$
$$
+{i (\l_1+i)(\l_2+i)^4(\l_3^2+\l_3(\l_1+i)+(\l_1+i)^2)\l_4^4
\over 2 (\l_2-\l_1+i)(\l_4-\l_3-i)}+
{i (\l_1+i)^4(\l_2+i)^4\l_4^4
\over 2 (\l_2-\l_1+i)(\l_3-\l_1-i)(\l_4-\l_3-i)}\sim
$$
$$
\sim\;-{i\over 2}{\l_2^4(\l_3+i)^4\l_4^3
\over(\l_2-\l_1)(\l_3-\l_1)}\;+\;
{i\over 2}{(\l_1+i)\l_2^4(\l_3^2+\l_3(\l_1+i)+(\l_1+i)^2)(\l_4+i)^4
\over(\l_2-\l_1)(\l_4-\l_3)}\;-\;
$$
$$
-\;{i\over 2}{\l_1^4\l_2^4(\l_4+i)^4
\over(\l_2-\l_1+i)(\l_3-\l_1)(\l_4-\l_3)}\;\sim\;
$$
$$
\sim\;-{i\over 2}{\l_2^4(\l_3+i)^4\l_4^3
\over(\l_2-\l_1)(\l_3-\l_1)}\;+\;
{i\over 2}{(\l_1+i)\l_2^4(\l_3^2+\l_3(\l_1+i)+(\l_1+i)^2)(\l_4+i)^4
\over(\l_2-\l_1)(\l_4-\l_3)}\;+\;
$$
$$
+{i (\l_1^3+\l_1^2(\l_2+i)+\l_1(\l_2+i)^2+(\l_2+i)^3)
\l_2^4(\l_4+i)^4
\over 2 (\l_3-\l_1)(\l_4-\l_3)}-
{i (\l_2+i)^4\l_2^4(\l_4+i)^4
\over 2 (\l_2-\l_1+i)(\l_3-\l_1)(\l_4-\l_3)}\sim
$$
$$
\sim\;-{i\over 2}{\l_2^4(\l_3+i)^4\l_4^3
\over(\l_2-\l_1)(\l_3-\l_1)}\;+\;
{i\over 2}{(\l_1+i)\l_2^4(\l_3^2+\l_3(\l_1+i)+(\l_1+i)^2)(\l_4+i)^4
\over(\l_2-\l_1)(\l_4-\l_3)}\;-\;
$$
\be
-\;{i\over 2}{(\l_2+i)^4\l_4^4
(\l_3^3+\l_3^2(\l_4+i)+\l_3(\l_4+i)^2+(\l_4+i)^3)
\over(\l_2-\l_1)(\l_3-\l_1)}\;+\;
{i\over 2}{\l_2^4(\l_2-i)^4(\l_4+i)^4
\over(\l_2-\l_1)(\l_3-\l_1)(\l_4-\l_3)}
\label{I44a}
\ee

Now we treat $I_6^{(4)}$ given by (\ref{I46}) as follows
$$
I_6^{(4)}\;\sim\;
i{(\l_1+i)(\l_2+i)\l_2^3\l_3^3(\l_4+i)^4
\over(\l_2-\l_1-i)(\l_3-\l_1-i)(\l_4-\l_3+i)}\;=\;
$$
$$
=\;i{(\l_1+i)(\l_2+i)(\l_2^2+\l_2(\l_1+i)+(\l_1+i)^2)\l_3^3(\l_4+i)^4
\over(\l_3-\l_1-i)(\l_4-\l_3+i)}\;+\;
$$
$$
+\;i{(\l_1+i)^4(\l_2+i)\l_3^3(\l_4+i)^4
\over(\l_2-\l_1-i)(\l_3-\l_1-i)(\l_4-\l_3+i)}\;=\;
$$
$$
=\;-i{(\l_2+i)(\l_2^2+\l_2(\l_1+i)+(\l_1+i)^2)\l_3^3(\l_4+i)^4
\over\l_4-\l_3+i}\;+\;
$$
$$
+\;i{(\l_2+i)(\l_2^2+\l_2(\l_1+i)+(\l_1+i)^2)\l_3^4(\l_4+i)^4
\over(\l_3-\l_1-i)(\l_4-\l_3+i)}\;+\;
i{(\l_1+i)^4(\l_2+i)\l_3^3(\l_4+i)^4
\over(\l_2-\l_1-i)(\l_3-\l_1-i)(\l_4-\l_3+i)}\;=\;
$$
$$
=\;i{(\l_2+i)(\l_2^2+\l_2(\l_1+i)+(\l_1+i)^2)\l_3^3\l_4^4
\over\l_4-\l_3}\;+\;
$$
$$
+\;i{(\l_2+i)(\l_2^2+\l_2(\l_1+i)+(\l_1+i)^2)(\l_3+i)^4(\l_4+i)^4
\over(\l_3-\l_1)(\l_4-\l_3)}\;+\;
i{\l_1^4(\l_2+i)\l_3^3\l_4^4
\over(\l_2-\l_1)(\l_3-\l_1)(\l_4-\l_3)}\;\sim\;
$$
$$
\sim\;-i{\l_1^3\l_2^4(\l_3+i)(\l_3^2+\l_3(\l_4+i)+(\l_4+i)^2)
\over\l_2-\l_1}\;-\;
$$
\be
-\;i{(\l_1+i)^4(\l_2+i)^4(\l_4^2+\l_4(\l_3+i)+(\l_3+i)^2)(\l_4+i)
\over(\l_2-\l_1)(\l_3-\l_1)}\;+\;
i{\l_1^4(\l_2+i)\l_3^3\l_4^4
\over(\l_2-\l_1)(\l_3-\l_1)(\l_4-\l_3)}
\label{I46a}
\ee

The last term $I_8^{(4)}$ (\ref{I48}) is more simple
$$
I_8^{(4)}\;\sim\;
-i{(\l_1+i)(\l_2+i)^4\l_3^3(\l_4+i)\l_4^3
\over(\l_2-\l_1+i)(\l_3-\l_1-i)(\l_4-\l_3+i)}\;=\;
$$
$$
=\;i{(\l_2+i)^4\l_3^3(\l_4+i)\l_4^3
\over(\l_2-\l_1+i)(\l_4-\l_3+i)}\;-\;
i{(\l_2+i)^4\l_3^4(\l_4+i)\l_4^3
\over(\l_2-\l_1+i)(\l_3-\l_1-i)(\l_4-\l_3+i)}\;=\;
$$
$$
=\;i{(\l_2+i)^4\l_3^3\l_4^3
\over\l_2-\l_1+i}\;+\;
i{(\l_2+i)^4\l_3^4\l_4^3
\over(\l_2-\l_1+i)(\l_4-\l_3+i)}\;-\;
i{(\l_2+i)^4\l_3^4(\l_4+i)\l_4^3
\over(\l_2-\l_1+i)(\l_3-\l_1-i)(\l_4-\l_3+i)}\;\sim\;
$$
\be
\sim\;i{\l_2^4(\l_3+i)^4\l_4^3
\over(\l_2-\l_1)(\l_4-\l_3)}\;-\;
i{\l_2^4(\l_3+i)^4(\l_4+i)\l_4^3
\over(\l_2-\l_1)(\l_3-\l_1)(\l_4-\l_3)}
\label{I48a}
\ee

Now we are prepared to perform the next our step. Namely, we will
gather all the six results (\ref{I41a}-\ref{I48a}) into the
form like (\ref{I123468})
$$
I_1^{(4)}\;+\;I_2^{(4)}\;+\;I_3^{(4)}\;+\;
I_4^{(4)}\;+\;I_6^{(4)}\;+\;I_8^{(4)}\;\sim\;
$$
\be
\sim\;{A\over (\l_2-\l_1)(\l_3-\l_1)(\l_4-\l_3)}\;+\;
{B\over (\l_2-\l_1)(\l_3-\l_1)}\;+\;
{C\over (\l_2-\l_1)(\l_4-\l_3)}\;+\;
{D\over \l_2-\l_1}
\label{I123468a}
\ee
where
$$
A\;=\;-\;{i\over 2}(\l_1+i)\l_2^4(\l_3+i)^4(\l_4+i)^3\;+\;
{i\over 2}\l_1^4(\l_2+i)\l_4^4(\l_4-i)^3\;+\;
{i\over 2}\l_1^4\l_2^3(\l_4+2i)(\l_4+i)^4\;+\;
$$
\be
+\;{i\over 2}\l_2^4(\l_2-i)(\l_4-i)^4\;+\;
i\l_1^4(\l_2+i)\l_3^3\l_4^4\;-\;
i\l_2^4(\l_3+i)^4(\l_4+i)\l_4^3
\label{A}
\ee
$$
B\;={i\over 2}(\l_1+i)^4(\l_2+i)^4(\l_3^2+\l_3(\l_4+i)+(\l_4+i)^2)
(\l_4+i)\;-\;i\l_2^4(\l_3+i)^4(\l_4+i)^3\;+\;
$$
$$
+\;{i\over 2}\l_1^4(\l_2+i)(\l_3^2+\l_3(\l_4+i)+(\l_4+i)^2)
(\l_4+i)\l_4^3\;+\;
i\l_1^4\l_2^3(\l_4+i)\l_4^3\;+\;
$$
\be
+\;i(\l_1+i)^4(\l_2+i)^4(\l_4^2+\l_4(\l_3+i)+(\l_3+i)^2)
(\l_4+i)
\label{B}
\ee
\be
C\;={i\over 2}(\l_1+i)\l_2^4(\l_3^2+\l_3(\l_1+i)+(\l_1+i)^2)
(\l_4+i)^4\;+\;i\l_2^4(\l_3+i)^4\l_4^3
\label{C}
\ee
$$
D\;={i\over 2}\l_1^3\l_2^4(\l_3^2+\l_3(\l_4+i)+(\l_4+i)^2)
(\l_4+i)\;-\;i\l_2^4\l_3^3(\l_4+i)^3\;-\;
$$
\be
-\;i\l_1^3\l_2^4(\l_3^2+\l_3(\l_4+i)+(\l_4+i)^2)
(\l_3+i)
\label{DD}
\ee

Now we want to get the "canonical" form (\ref{T_n^c}) for the
expression (\ref{I123468a}).
To do this we actively used the program {\it MATHEMATICA}
because the calculations are straightforward but become more cumbersome.
Let us outline our further actions.

It is more convenient to start with the first term in the r.h.s.
of the formula (\ref{I123468a})
\be
{A\over (\l_2-\l_1)(\l_3-\l_1)(\l_4-\l_3)}
\label{Aa}
\ee
We shall use the fact that the denominator is antisymmetric under
the substitution 
\beq
\l_1\leftrightarrow\l_3\nonumber\\
\l_2\leftrightarrow\l_4\nonumber
\eeq
Since $A$ is a polynomial given by (\ref{A}) 
then the simplification procedure of the term (\ref{Aa})
is as follows. If in the expression (\ref{A}) one faces 
a monomial 
\be
\l_1^{i_1}\l_2^{i_2}\l_3^{i_3}\l_4^{i_4}
\label{mon1234}
\ee
where without loss of generality $i_4\ge i_2$ 
one can apply the evident identity
\be
\l_4^{i_4}\;=\;\l_4^{i_2}\l_3^{i_4-i_2}\;+\;
\cases{(\l_4-\l_3)\sum_{k=0}^{i_4-i_2-1}
\l_4^{i_2+k}\l_3^{i_4-i_2-1-k}, &
$i_4>i_2;$\cr
0 \quad,& $i_4=i_2$\cr}
\label{l4i4}
\ee
Therefore if $i_4>i_2$ then
\be
{\l_1^{i_1}\l_2^{i_2}\l_3^{i_3}\l_4^{i_4}\over
(\l_2-\l_1)(\l_3-\l_1)(\l_4-\l_3)}\;=\;
{\l_1^{i_1}\l_2^{i_2}\l_3^{i_4-i_2+i_3}\l_4^{i_2}\over
(\l_2-\l_1)(\l_3-\l_1)(\l_4-\l_3)}\;+\;
\sum_{k=0}^{i_4-i_2-1}
{\l_1^{i_1}\l_2^{i_2}\l_3^{i_4-i_2+i_3-1-k}\l_4^{i_2+k}\over
(\l_2-\l_1)(\l_3-\l_1)}
\label{l4mon}
\ee
Let us note that the second term in (\ref{l4mon})
gives rise into the second term $"B"$ in the r.h.s. 
of the formula (\ref{I123468a}).
If $i_4=i_2$ then only the first term in (\ref{l4mon}) survives.
The first term in (\ref{l4mon}) is symmetric under the 
transposition $\l_2\leftrightarrow\l_4$. If $i_1+i_2=i_3+i_4$
then it is also symmetric under the transposition
$\l_1\leftrightarrow\l_3$ and the first term is "weakly"
equivalent to zero according to the formula (\ref{int}).
If $i_1+i_2<i_3+i_4$ the following
"weak" equality is valid
$$
{\l_1^{i_1}\l_2^{i_2}\l_3^{i_4-i_2+i_3}\l_4^{i_2}\over
(\l_2-\l_1)(\l_3-\l_1)(\l_4-\l_3)}\;\sim\;
{1\over 2}
{(\l_1^{i_1}\l_3^{i_3+i_4-i_2}-\l_3^{i_1}\l_1^{i_3+i_4-i_2})
\l_2^{i_2}\l_4^{i_2}\over
(\l_2-\l_1)(\l_3-\l_1)(\l_4-\l_3)}\;=\;
$$
\be
=\;{1\over 2}\sum_{k=0}^{i_3+i_4-i_2-i_1-1}
{\l_1^{i_1+k}\l_2^{i_2}\l_3^{i_3+i_4-i_2-1-k}\l_4^{i_2}\over
(\l_2-\l_1)(\l_4-\l_3)}
\label{l31}
\ee
For the case $i_1+i_2>i_3+i_4$ the sum in (\ref{l31}) should be
substituted by
\be
-{1\over 2}\sum_{k=0}^{i_1+i_2-i_3-i_4-1}
{\l_1^{i_1-1-k}\l_2^{i_2}\l_3^{i_3+i_4-i_2+k}\l_4^{i_2}\over
(\l_2-\l_1)(\l_4-\l_3)}
\label{l31a}
\ee
In both cases, namely, if $i_1+i_2\ne i_3+i_4$ the sum in (\ref{l31})
or (\ref{l31a}) gives rise into the third term "C" in the r.h.s.
of (\ref{I123468a}).

Performing this procedure for all monomials of the form (\ref{mon1234})
participating in the polynomial $A$ given by the formula (\ref{A}) 
one can arrive at the formula
\be
I_1^{(4)}\;+\;I_2^{(4)}\;+\;I_3^{(4)}\;+\;
I_4^{(4)}\;+\;I_6^{(4)}\;+\;I_8^{(4)}\;\sim\;
{B'\over (\l_2-\l_1)(\l_3-\l_1)}\;+\;
{C'\over (\l_2-\l_1)(\l_4-\l_3)}\;+\;
{D\over \l_2-\l_1}
\label{I123468b}
\ee
with some other polynomials $B'$ and $C'$.

Due to the $2\leftrightarrow 3$ symmetry of the denominator
of the first term in (\ref{I123468b}) one can treat any
monomial $\l_1^{j_1}\l_2^{j_2}\l_3^{j_3}\l_4^{j_4}$ participating
in $B'$ as follows
\be
{\l_1^{j_1}\l_2^{j_2}\l_3^{j_3}\l_4^{j_4}\over
(\l_2-\l_1)(\l_3-\l_1)}\;\sim\;
\sum_{k=0}^{j_3-j_2-1}
{\l_1^{j_1}\l_2^{j_2+k}\l_3^{j_3-k-1}\l_4^{j_4}\over
\l_2-\l_1}
\label{monB'}
\ee
where without loss of generality it is implied that $j_3>j_2$
because if $j_2=j_3$ then due to the $2\leftrightarrow 3$ symmetry
and the formula (\ref{int}) this term would make zero
contribution. So the r.h.s. of the (\ref{monB'}) gives rise
into the third term in (\ref{I123468b}).
Using this one can treat the whole first term in (\ref{I123468b}).

For any monomial $\l_1^{k_1}\l_2^{k_2}\l_3^{k_3}\l_4^{k_4}$ 
participating in $C'$ of the second term of the expression 
(\ref{I123468b}) one can write 
\be
{\l_1^{k_1}\l_2^{k_2}\l_3^{k_3}\l_4^{k_4}
\over (\l_2-\l_1)(\l_4-\l_3)}\;=\;
\sum_{l=0}^{k_2-1}
{\l_1^{k_1+l}\l_2^{k_2}\l_3^{k_3-1-l}\l_4^{k_4}
\over \l_4-\l_3}\;+\;
\sum_{l=0}^{k_4-1}
{\l_1^{k_1+k_2}\l_3^{k_3+l}\l_4^{k_4-1-l}
\over \l_2-\l_1}\;+\;
{\l_1^{k_1+k_2}\l_3^{k_3+k_4}\over (\l_2-\l_1)(\l_4-\l_3)}\;\sim\;
\label{l12a}
\ee
\be
\sim\;\sum_{l=0}^{k_2-1}
{\l_1^{k_1}\l_2^{k_4}\l_3^{k_1+l}\l_4^{k_2-1-l}
\over \l_2-\l_1}\;+\;
\sum_{l=0}^{k_4-1}
{\l_1^{k_1+k_2}\l_3^{k_3+l}\l_4^{k_4-1-l}
\over \l_2-\l_1}\;+\;
{\l_1^{k_1+k_2}\l_3^{k_3+k_4}\over (\l_2-\l_1)(\l_4-\l_3)}
\label{l12b}
\ee
where for the first sum of (\ref{l12a}) we have applied
transformation 
\beq
\l_1\leftrightarrow\l_3\nonumber\\
\l_2\leftrightarrow\l_4\nonumber
\eeq
So the both the first term and the second term in (\ref{l12b})
give rise into the third "D" term in (\ref{I123468b}) while the
third term in (\ref{l12b}) gives rise to the second term "C"
in (\ref{I123468b}).
Proceeding this way one can treat the whole expression
$$
{C'\over (\l_2-\l_1)(\l_4-\l_3)}.
$$

As a result of performing this scheme one can arrive at the
expression
$$
I_1^{(4)}\;+\;I_2^{(4)}\;+\;I_3^{(4)}\;+\;
I_4^{(4)}\;+\;I_6^{(4)}\;+\;I_8^{(4)}\;\sim\;
$$
\be
\sim\;{C''\over (\l_2-\l_1)(\l_4-\l_3)}\;+\;
{D'\over \l_2-\l_1}
\label{I123468c}
\ee
where $C''$ is a polynomial of two variable $\l_1$ and $\l_3$.

Now with the help of the identity 
\be
\l_1^{i_1}\l_2^{i_2}\;=\;\l_1^{i_1+i_2}\;+\;
\cases{(\l_2-\l_1)\sum_{k=0}^{i_2-1}
\l_1^{i_1+k}\l_2^{i_2-1-k}, & $i_2>0;$\cr
0 \quad,& $i_2=0$\cr}
\label{l12}
\ee
one can reduce the second term in (\ref{I123468c}) to the form
$$
{D'\over \l_2-\l_1}\;\sim\;{D''\over \l_2-\l_1}\;+\;E
$$
where $D''$ is a polynomial of $\l_1,\l_3$ and $\l_4$ and $E$ is
some polynomial.

What is left now is to reduce the power of the polynomials 
$C'',D''$ and $E$ with the help of the formulae (\ref{l^m}),
(\ref{cor1}), (\ref{cor2}) or (\ref{l^3}-\ref{l^6}). We should
also use the fact that if we do the substitution $\l_j\rightarrow
x_j-i/2$ there is a restriction that the function should be 
even under the transformation $\{x_1,x_2,x_3,x_4\}\rightarrow
\{-x_1,-x_2,-x_3,-x_4\}$ according to the formulae (\ref{tildeU}),
(\ref{tildeF}) and (\ref{P_ja}). 

As a result of all these actions described above we get the "canonical"
form for the sum 
\be
I_1^{(4)}\;+\;I_2^{(4)}\;+\;I_3^{(4)}\;+\;
I_4^{(4)}\;+\;I_6^{(4)}\;+\;I_8^{(4)}\;\sim\;
{C'''\over (\l_2-\l_1)(\l_4-\l_3)}\;+\;
{D'''\over \l_2-\l_1}\;+\;E'
\label{I123468d}
\ee
where
\be
C'''\;=\;2\l_1^2\l_3^2\;+\;4i\l_1\l_3^2\;-\;{3\over 2}\l_3^2
\;-\;{3\over 2}\l_1\l_3\;-\;i\l_3\;+\;{1\over 5}
\label{C'''}
\ee
$$
D'''\;=\;
\l_1^2\,(22\l_3^2\l_4^3\;+\;22i\l_3\l_4^3\;-\;{29\over 2}\l_4^3\;+\;
19\l_3\l_4^2\;-\;{5\over 4}i\l_4^2)\;+\;
$$
$$
+\;\l_1\,(22i\l_3^2\l_4^3\;-\;{47\over 2}\l_3\l_4^3\;-\;
{61\over 4}i\l_4^3\;+\;{67\over 4}i\l_3\l_4^2\;-\;
{3\over 2}\l_4^2\;+\;{23\over 4}i\l_4)\;-\;
$$
\be
-\;{88\over 5}\l_3^2\l_4^3\;-\;{367\over 20}i\l_3\l_4^3\;+\;
{427\over 40}\l_4^3\;-\;{303\over 40}\l_3\l_4^2\;+\;
{37\over 40}i\l_4^2\;-\;{97\over 20}\l_4)
\label{D'''}
\ee
\be
E'\;=\;-\;{57\over 10}\l_2\l_3^2\l_4^3
\label{E'}
\ee

Now we have to sum up the four contributions (\ref{T4b}), (\ref{I579}),
(\ref{I410a}) and (\ref{I123468d}) and get the "canonical" form for
$T_4(\l_1,\l_2,\l_3,\l_4)$
\be
T_4(\l_1,\l_2,\l_3,\l_4)\;\sim\;
P_0^{(4)}\;+\;{P_1^{(4)}\over\l_2-\l_1}\;+\;
{P_2^{(4)}\over(\l_2-\l_1)(\l_4-\l_3)}
\label{T_4^c}
\ee
where 
\be
P_0^{(4)}\;=\;-{34\over 5}\l_2\l_3^2\l_4^3
\label{P40}
\ee
$$
P_1^{(4)}\;=\;
\l_1^2\,(30\,\l_3^2\l_4^3\;+\;30\,i\l_3\l_4^3\;-\;16\,\l_4^3\;+\;
18\,\l_3\l_4^2\;+\;8\,\l_4)\;+\;
$$
$$
+\;\l_1\,(30\,i\l_3^2\l_4^3\;+\;30\,\l_3\l_4^3\;-\;
16\,i\l_4^3\;+\;18\,i\l_3\l_4^2\;-\;
4\,\l_4^2\;+\;4\,i\l_4)\;-\;
$$
\be
-\;20\,\l_3^2\l_4^3\;-\;20\,i\l_3\l_4^3\;+\;
{54\over 5}\l_4^3\;-\;{42\over 5}\l_3\l_4^2\;-\;
{43\over 10}i\l_4
\label{P41}
\ee
\be
P_2^{(4)}\;=\;2\l_1^2\l_3^2\;+\;4i\l_1\l_3^2\;-\;{3\over 2}\l_3^2
\;-\;{3\over 2}\l_1\l_3\;-\;i\l_3\;+\;{1\over 5}
\label{P42}
\ee

Let us note that in terms of the real variables $x_j$
the polynomials $\tilde P_j^{(4)}$ (see (\ref{P_ja}))
look a little bit simpler
\be
\tilde P_0^{(4)}\;=\;-{34\over 5}x_2x_3^2x_4^3
\label{P40a}
\ee
$$
\tilde P_1^{(4)}\;=\;x_1^2\;(30\,x_3^2x_4^3\;-\;{17\over 2}x_4^3
\;+\;{81\over 2}x_3x_4^2\;+\;{79\over 8}x_4)\;-\;4\,x_1x_4^2\;-\;
$$
\be
-\;{25\over 2}x_3^2x_4^3
\;+\;{147\over 40}x_4^3\;-\;{531\over 40}x_3x_4^2\;-\;
{653\over 160}x_4
\label{P41a}
\ee
\be
\tilde P_2^{(4)}\;=\;2\,x_1^2x_3^2\;+\;{1\over 2}x_1x_3\;-\;
{1\over 2}x_3^2\;+\;{3\over 40}
\label{P42a}
\ee
So, the function 
\be
\tilde T_4^c(x_1,x_2,x_3,x_4)\;=\;
\tilde P_0^{(4)}\;+\;{\tilde P_1^{(4)}\over x_2-x_1}
\;+\;{\tilde P_2^{(4)}\over (x_2-x_1)(x_4-x_3)}
\label{tildeT_4^c}
\ee
is even i.e. 
$$
\tilde T_4^c(-x_1,-x_2,-x_3,-x_4)\;=\;\tilde T_4^c(x_1,x_2,x_3,x_4)
$$
as it should be according to the formula (\ref{tildeT_n^ca}) with $n=4$.

Now let us start the second step of our general scheme, namely,
the performing of the integration of the "canonical" form 
(\ref{T_4^c})
\be
P(4)\;=\;\prod_{j=1}^4\int_{C_{-1/2}}{d\l_j\over 2\pi i}\>
U_4(\l_1,\l_2,\l_3,\l_4)\>T_4(\l_1,\l_2,\l_3,\l_4)\;=\;
J_0^{(4)}\;+\;J_1^{(4)}\;+\;J_2^{(4)}
\label{P4a}
\ee
where
\be
J_0^{(4)}=\prod_{j=1}^4\int_{C_{-1/2}}{d\l_j\over 2\pi i}\>
U_4(\l_1,\l_2,\l_3,\l_4)\>P_0^{(4)}
\label{J40}
\ee
\be
J_1^{(4)}=\prod_{j=1}^4\int_{C_{-1/2}}{d\l_j\over 2\pi i}\>
U_4(\l_1,\l_2,\l_3,\l_4)\>
{P_1^{(4)}\over\l_2-\l_1}
\label{J41}
\ee
\be
J_1^{(4)}=\prod_{j=1}^4\int_{C_{-1/2}}{d\l_j\over 2\pi i}\>
U_4(\l_1,\l_2,\l_3,\l_4)\>
{P_2^{(4)}\over(\l_2-\l_1)(\l_4-\l_3)}
\label{J42}
\ee
Using the formulae (\ref{J40}-\ref{J42}),
(\ref{P40}-\ref{P42}),(\ref{F}),(\ref{Da}),
(\ref{ro5}) we get
$$
J_0^{(4)}\;=\;\prod_{j=1}^4\int_{C_{-1/2}}{d\l_j\over 2\pi i}\>
U_4(\l_1,\l_2,\l_3,\l_4)\>(-{34\over 5})\l_2\l_3^2\l_4^3\;=\;
$$
$$
=\;D^{(4)}\sum_{l_1=0}^{\infty}(-1)^{l_1}
\sum_{l_2=0}^{\infty}(-1)^{l_2}\sum_{l_3=0}^{\infty}(-1)^{l_3}
\sum_{l_4=0}^{\infty}(-1)^{l_4}
(-{34\over 5})(i l_2+\e_2)(i l_3+\e_3)^2(i l_4+\e_4)^2\;=
$$
$$
=\;D^{(4)}\sum_{l_1=0}^{\infty}(-1)^{l_1}
\sum_{l_2=0}^{\infty}(-1)^{l_2}\sum_{l_3=0}^{\infty}(-1)^{l_3}
\sum_{l_4=0}^{\infty}(-1)^{l_4}
(-{34\over 5})\e_2\e_3^2\e_4^3\;=\;
$$
\be
=\;{1\over 12}\prod_{0\le k<j\le 4}
{\Bigl({\partial\over\partial\e_k}-
{\partial\over\partial\e_j}\Bigr)}_
{\vec\e\rightarrow 0}
(-{34\over 5})\e_2\e_3^2\e_4^3\;=\;-{17\over 40}
\label{J40a}
\ee

Restoring for convenience the dependence of the polynomials
$P_1^{(4)}$ and $P_2^{(4)}$ on $\l$-s as in the formula (\ref{P_j})
we can get for the term $J_1^{(4)}$ 
$$
J_1^{(4)}\;=\;\prod_{j=1}^4\int_{C_{-1/2}}{d\l_j\over 2\pi i}\>
U_4(\l_1,\l_2,\l_3,\l_4)\>{P_1^{(4)}(\l_1|\l_3,\l_4)
\over \l_2-\l_1}\;=\;
$$
$$
=\;D^{(4)}\sum_{l_1=1}^{\infty}(-1)^{l_1}
\sum_{l_2=0}^{\infty}(-1)^{l_2}\sum_{l_3=1}^{\infty}(-1)^{l_3}
\sum_{l_4=0}^{\infty}(-1)^{l_4}
{P_1^{(4)}(-il_1+\e_1|-il_3+\e_3,il_4+\e_4)
\over il_2+il_1+\e_2-\e_1}\;=\;
$$
$$
=\;-\;i D^{(4)}\int_0^1{ds\over s}\sum_{l_1=1}^{\infty}(-1)^{l_1}
\sum_{l_2=0}^{\infty}(-1)^{l_2}\sum_{l_3=1}^{\infty}(-1)^{l_3}
\sum_{l_4=0}^{\infty}(-1)^{l_4}
$$
$$
P_1^{(4)}(-il_1+\e_1|-il_3+\e_3,il_4+\e_4)
s^{l_1+l_2+i(\e_1-\e_2)}\;=
$$
$$
=\;\int_0^1 ds \Bigl(-{15\over 2}{(s-1)\over(1+s)^3}\;+\;
{3\over 4}{(7-26s+7s^2)\over(1+s)^4}\ln{s}\;+\;
{21\over 8}{(s-1)\over(1+s)^3}\ln^2{s}\;-\;
{7\over 10}{(2-s+2s^2)\over(1+s)^4}\ln^3{s}\;+\;
$$
\be
+\;{5\over 48}{(s-1)\over(1+s)^3}\ln^4{s}\;+\;
{1\over 240}{(1+22s+s^2)\over(1+s)^4}\ln^5{s}\Bigr)\;=\;
{5\over 8}\;-\;2\ln{2}\;+\;{61\over 20}\zeta(3)\;-\;
{65\over 32}\zeta(5)
\label{J41a}
\ee

The last term $J_2^{(4)}$ can be calculated as follows
$$
J_2^{(4)}\;=\;\prod_{j=1}^4\int_{C_{-1/2}}{d\l_j\over 2\pi i}\>
U_4(\l_1,\l_2,\l_3,\l_4)\>{P_2^{(4)}(\l_1,\l_3)
\over (\l_2-\l_1)(\l_4-\l_3)}\;=\;
$$
$$
=\;D^{(4)}\sum_{l_1=1}^{\infty}(-1)^{l_1}
\sum_{l_2=0}^{\infty}(-1)^{l_2}\sum_{l_3=1}^{\infty}(-1)^{l_3}
\sum_{l_4=0}^{\infty}(-1)^{l_4}
{P_2^{(4)}(-il_1+\e_1,-il_3+\e_3)
\over (il_2+il_1+\e_2-\e_1)(il_4+il_3+\e_4-\e_3)}\;=\;
$$
$$
=\;-\;D^{(4)}\int_0^1{ds\over s}
\int_0^1{dt\over t}\sum_{l_1=1}^{\infty}(-1)^{l_1}
\sum_{l_2=0}^{\infty}(-1)^{l_2}\sum_{l_3=1}^{\infty}(-1)^{l_3}
\sum_{l_4=0}^{\infty}(-1)^{l_4}
$$
$$
P_2^{(4)}(-il_1+\e_1,-il_3+\e_3)
s^{l_1+l_2+i(\e_1-\e_2)}t^{l_3+l_4+i(\e_3-\e_4)}\;=\;
$$
$$
=\;\int_0^1 ds\int_0^1 dt 
\Bigl\{4{\ln{s}(\ln{s}-\ln{t})\over(1+s)^2(1+t)^2}\;+\;
(-{4\over 3}(-3-s+t+3st)\ln^3{s}\;+\;4(-1+3s-3t+st)\ln^2{s}\ln{t}\;-\;
$$
$$
-\;{3\over 4}(3-13s+3t+3st)\ln^2{s}\ln^2{t}\;+\;
{5\over 12}(3-5s-5t+3st)\ln^4{s}){1\over(1+s)^3(1+t)^3}\;-\;
$$
$$
-\;{1\over 6}{(-1+5s-2s^2+2t-5st+s^2t)\over (1+s)^4(1+t)^3}
\ln{s}(\ln^2{s}-\ln^2{t})
(\ln^2{s}-5\ln^2{t})\;+\;
$$
$$
+\;({1\over 3}(3+30s-5s^2-34t-4st-34s^2t-5t^2+30st^2+3s^2t^2)
\ln^3{s}\ln{t}\;+\;
$$
$$
+\;{1\over 30}(4-17s+9s^2-17t+66st-17s^2t+9t^2-17st^2+4s^2t^2)
\ln^5{s}\ln{t}\;-\;
$$
$$
-\;{1\over 15}(2-31s+7s^2+14t+33st-21s^2t+2t^2+4st^2+2s^2t^2)
\ln^3{s}\ln^3{t}){1\over(1+s)^4(1+t)^4}\Bigl\}
$$
Taking the integral we come to an answer for $J_2^{(4)}$
\be
J_2^{(4)}\;=\;-{1\over 6}\zeta(3)\;-\;{11\over 6}\zeta(3)\ln{2}\;-\;
{51\over 80}\zeta(3)^2\;-\;{25\over 96}\zeta(5)\;+\;
{85\over 24}\zeta(5)\ln{2}
\label{J42a}
\ee

Summing up all the three results (\ref{J40a}), (\ref{J41a}) and
(\ref{J42a}) we finally get our main formula (\ref{P4}) 
$$
P(4)\;=\;J_0^{(4)}\;+\;J_1^{(4)}\;+\;J_2^{(4)}\;=\;
$$
\be
{1\over 5}\; -\; 2\ln{2} \; + \; {173\over 60}\,\zeta(3)
\; -\; {11\over 6} \,\zeta(3)\, \ln{2}\;-\; {51\over 80} \, \zeta^2(3)
 \;-\;
 {55\over 24}\,\zeta(5)
\; + \; {85\over 24}\,\zeta(5)\,  \ln{2}
\label{P4b}
\ee

\end{document}